\documentclass[10pt,aps,pra,twocolumn,showpacs,showkeys,superscriptaddress,longbibliography]{revtex4-2}

\usepackage[table]{xcolor}
\usepackage{times}
\usepackage{mathtools}
\usepackage{dsfont} % for "openone" and in complex set of numbers Z
\usepackage{siunitx}
\usepackage{graphicx}
\usepackage{listings}
\usepackage{sourcecodepro}
\usepackage{relsize}

\newcommand{\tabdesc}[1]{
  \parbox[t]{11.105cm}{\raggedright #1\strut}
}

\definecolor{safeOrange}{RGB}{230, 159, 0}      % Replaces mygold/red
\definecolor{safeSkyBlue}{RGB}{86, 180, 233}    % Replaces mycyan
\definecolor{safeGreen}{RGB}{0, 158, 115}       % Replaces mygreen
\definecolor{safeYellow}{RGB}{240, 228, 66}     % High-contrast highlight
\definecolor{safeBlue}{RGB}{0, 114, 178}        % Replaces myblue
\definecolor{safeVermillion}{RGB}{213, 94, 0}   % Replaces myred
\definecolor{safePurple}{RGB}{204, 121, 167}    % Replaces mypurple
\definecolor{safeBlack}{RGB}{0, 0, 0}           % Anchor colour
\definecolor{myDarkGrey}{RGB}{128, 128, 128}
\definecolor{myLightGrey}{RGB}{224, 224, 224}

\renewcommand{\UrlFont}{\ttfamily\footnotesize\color{black}}
\newcommand{\simpson}[1]{{\ttfamily\footnotesize\color{black}\detokenize{#1}}} %$
\newcommand{\CC}{C\nolinebreak[4]\hspace{-.05em}\raisebox{.2ex}{\smaller\bf ++}}

\lstdefinelanguage{SIMPSON}{
    language     = tcl,       % Inherit everything from tcl
    morekeywords=[2]{ % Tcl_ObjCmdProc
            rand_shape, % stochastics
            isotopes, resfreq, dist2dip ,dip2dist, % spinsys
            internalsimpson, crystallites, % simpson
            load_shape, save_shape, free_shape, free_all_shapes, shape_len, shape_index, list2shape, shape2list, shape_join, shape_dup, shape_ampl, shape_energy, shape_manipulate, shape_create, shape_index_set, % rfshapes
            filter, pulse, reset, pulseid, delay, maxdt, acq, evolve, prop, store, offset, select, matrixoper, getinteractions, currenttime, rotorangle, avgham_static, pulse_shaped, eulerangles, zgrad_pulse_shaped, pulse_and_zgrad_shaped ,acq_block, acq_modulus, pulse_shaped_rotormodulated, test_function, % pulse
            oc_evaluate_grad, oc_grad_shapes, oc_gradmode, TF_line, % optimize
            create_distortion_operator, distort_shape, reconstruct_gradient, free_distortion_operator, % distortions
            oc_acq_hermit, oc_acq_nonhermit, oc_acq_prop, oc_grad_add_energy_penalty, oc_rflimit_penalty, oc_grad_add_rflimit_penalty, % OCroutines
        }, 
    morekeywords=[3]{ %Tcl_CmdProc
            gamma, %spinsys
            turnon, turnoff, matrix, % pulse
            oc_optimize, oc_optimize_phase, % optimize
            fabs, fload, floaddata, funload, fsave, fphase, fscale, fdup, fdupzero, frms, frealrms, fautoscale, fextract, fzero, fzerofill, frev, fget, fset, fcreate, fft, faddpeaks, fint, fssbint, fadd, fsub, fcopy, faddlb, findex, fx, ffindpeaks, fmaxheight, fsetindex, fplot2d, faddtriangle, freconstruct, fcompare, fnewnp, read, fft1d, ftranspose, fsplit, fbc, fsmooth, % ftools
            zgrad_shape_create, free_zgrad, load_zgrad, zgrad2list, % B0inhom
        },
    morekeywords=[4]{ %proc
            markvars, savestate, signalhandler, setvar, ssSubstExpr, ssSetValues, ssSetSpinsys, spinsys_resolve, spinsys, cmp_length, par, fsimpson, main, pulseq, % proc, procedures
            fexpr, % ftools proc
            csapar, csaprinc, gval2gtensor, gtensor2gval, hyperfine2hypAB, hypAB2hyperfine, putmatrix, contourplot, 2dplot, 3dplot, simview, % misc proc
            shape2fid, shape2varian, shape2bruker, % rfshapes proc
            relax, % relax proc
            putmatrix, par, spinsys, % slave proc
            target_function, gradient,  % Tcl_EvalEx
        },
    morekeywords=[4]{\$par,\$spinsys},
    morekeywords=[5]{ % proc spinsys
            channels, nuclei, dipole, quadrupole, shift, jcoupling, mixing, dipole_ave, gtensor, hyperfine, heisenberg, edipole, edipole_ave, quadrupole_x_shift, quadrupole_x_dipole,
        },
    morekeywords=[6]{ % par proc
            proton_frequency, magnetic_field, spin_rate, sw, sw1, np, ni, method, rotor_angle, gamma_angles, fixed_rep, real_spec, block_diag, detect_operator, crystal_file, start_operator, name, verbose, various, variable, pulse_sequence, conjugate_fid, dipole_check, gamma_zero, use_cluster, new_cluster, cluster_port, inner_rotor_angle, outer_rotor_angle, inner_spin_rate, outer_spin_rate, dor, string, oc_tol_cg, oc_tol_ls, oc_mnbrak_step, oc_max_iter, oc_cutoff, oc_cutoff_iter, oc_var_save_iter, oc_var_save_proc, oc_cg_min_step, oc_max_brack_eval, oc_max_brent_eval, oc_silent, oc_verbose, oc_grad_level, oc_method, oc_optm_method, oc_ctrl_method, oc_state_store_dim, oc_adapt_iter, rfprof_file, use_3_angle_set, acq_adjoint, zprofile, zvals, relax, prop_method, split_order, use_sparse, num_cores, averaging_file, points_per_cycle, ED_symmetry, oc_pulse_nreps, oc_lbfgs_eps, oc_lbfgs_tol_ls, oc_lbfgs_max_ls_eval, oc_lbfgs_m, sparsity, sparse_tol, maxfulldim, maxdimdiagonalize, do_avg, rfmap, 
        },
    morecomment=[l]{\#},
}

\lstdefinestyle{SIMPSON-coloured}{
    language = SIMPSON,
    basicstyle=\ttfamily\scriptsize\color{black!75},
    identifierstyle=\ttfamily\scriptsize\color{black!75},
    keywordstyle=\color{safeVermillion},
    keywordstyle=[2]\color{safeGreen},
    keywordstyle=[3]\color{safeGreen},
    keywordstyle=[4]\color{safeBlue},
    keywordstyle=[5]\color{safePurple},
    keywordstyle=[6]\color{safePurple},
    stringstyle=\color{safeOrange},
    commentstyle=\color{myDarkGrey},
    literate={\%.0f}{{\textcolor{safeOrange}{\%.0f}}}4, 
}
\apptocmd{\sloppy}{\hbadness 10000\relax}{}{}
\graphicspath{{./figures/}}
\usepackage{orcidlink}
\usepackage{hyperref}

\begin{document}

\title{Extending numerical simulations in SIMPSON: Electron paramagnetic resonance, dynamic nuclear polarisation, propagator splitting, pulse transients, and quadrupolar cross terms}

\author{David L. Goodwin\,\orcidlink{0000-0001-9423-1106}}
\email[]{david.goodwin@inano.au.dk}
\affiliation{Interdisciplinary Nanoscience Center (iNANO) and Department of Chemistry, Aarhus University, Gustav Wieds Vej 14, DK-8000 Aarhus C, Denmark}

\author{Jos\'{e} P. Carvalho\,\orcidlink{0000-0001-5648-4612}}
\affiliation{Interdisciplinary Nanoscience Center (iNANO) and Department of Chemistry, Aarhus University, Gustav Wieds Vej 14, DK-8000 Aarhus C, Denmark}

\author{Anders B. Nielsen\,\orcidlink{0000-0002-3937-1170}}
\affiliation{Interdisciplinary Nanoscience Center (iNANO) and Department of Chemistry, Aarhus University, Gustav Wieds Vej 14, DK-8000 Aarhus C, Denmark}

\author{Nino Wili\,\orcidlink{0000-0003-4890-3842}}
\affiliation{Interdisciplinary Nanoscience Center (iNANO) and Department of Chemistry, Aarhus University, Gustav Wieds Vej 14, DK-8000 Aarhus C, Denmark}

\author{Thomas Vosegaard\,\orcidlink{0000-0001-5414-4550}}
\email[]{tv@chem.au.dk}
\affiliation{Interdisciplinary Nanoscience Center (iNANO) and Department of Chemistry, Aarhus University, Gustav Wieds Vej 14, DK-8000 Aarhus C, Denmark}

\author{Zden{\v{e}}k To{\v{s}}ner\,\orcidlink{0000-0003-2741-9154}}
\email[]{zdenek.tosner@natur.cuni.cz}
\affiliation{Department of Chemistry, Faculty of Science, Charles University in Prague, Hlavova 8, CZ-128 43, Czech Republic}

\author{Niels Chr. Nielsen\,\orcidlink{0000-0003-2978-4366}}
\email[]{ncn@chem.au.dk}
\affiliation{Interdisciplinary Nanoscience Center (iNANO) and Department of Chemistry, Aarhus University, Gustav Wieds Vej 14, DK-8000 Aarhus C, Denmark}

\date{February 2026}

\begin{abstract}
Aimed at the simulation, design, and interpretation of advanced pulse experiments crossing the boundaries between nuclear magnetic resonance (NMR) and electron paramagnetic resonance (EPR), including the rapidly emerging, hybrid discipline of pulsed dynamic nuclear polarisation (DNP), we present a host of novel features in the widely used SIMPSON software package addressing these aspects. Along with this come new features for advanced pulse sequence evaluation in terms of propagator splitting, high-order spin operator cross terms, and pulse phase transients. These fundamental new tools are introduced in a C++-based next generation of the SIMPSON software, which improves calculations speed in some aspects, is better prepared for further developments, and facilitates easier community contributions to the open-source software package.
\end{abstract}


%\pacs{--.--.--, --.--.--, --.--.--, --.--.--}

\maketitle

\section{Introduction}

In addition to the significant developments in magnetic resonance instrumentation and advances in fundamental theoretical descriptions, versatile numerical simulation tools presented in advanced software packages have played an invaluable role in development, understanding, and application of methods in magnetic resonance. This is pertinent to nuclear magnetic resonance (\textsc{nmr}) and nuclear magnetic resonance imaging (\textsc{mri}). However, with the development of fast waveform generators and powerful pulsed microwave amplifiers, this also applies increasingly to electron paramagnetic resonance (\textsc{epr}; or electron spin resonance, \textsc{esr}) and dynamic nuclear polarisation (\textsc{dnp}). Complementing a large variety of stand-alone programmes, specialised aspects of numerical simulations, and data processing software, several more general software packages have been developed for numerical simulation of advanced \textsc{nmr} pulse sequences: \textsc{antiope} \cite{deBouregas1992}; Gamma \cite{Smith1994}; \textsc{simpson} \cite{Bak2000}; wSolids \cite{Eichele2001}; DMfit \cite{Massiot2002}; \textsc{spinevolution} \cite{Veshtort2006}; Spinach \cite{Hogben2011}; SpinDynamica \cite{Bengs2018}; MolSpin \cite{Nielsen2019}; ssNake \cite{vanMeerten2019}; \textsc{mrs}imulator \cite{Srivastava2024}; Sleepy \cite{Smith2025}. Each software package has its particular features and strengths, which are not addressed specifically in this work, except by noting that the software packages \textsc{simpson} (simulation package for solid-state \textsc{nmr} spectroscopy) and Spinach probably feature the most general aspects of numerical simulation in solid-state \textsc{nmr}. Likewise, software has been presented for the simulation of \textsc{epr} spectra and pulse sequences; regarding pulse sequences in particular, attention should be drawn to the general and widely used software package EasySpin \cite{Stoll2006,Stoll2014} and components in Spinach \cite{Hogben2011}. Both of these are also relevant for the simulation of \textsc{dnp} experiments, as are Gamma \cite{Smith1994} and \textsc{spinevolution} \cite{Veshtort2006}. Furthermore, more specific software for \textsc{dnp} has been introduced recently, including \textsc{dnps}oup \cite{Yang2022}. Our focus in this work is \textsc{simpson} which, since its introduction to the \textsc{nmr} community in 2000, has been one of the most widely used and extensively cited software packages for solid-state \textsc{nmr}. Since the first publication and release as open-source software \cite{Bak2000}, numerous papers have been presented with each introducing new features: parallelization \cite{Tosner2014}; optimal control \cite{Kehlet2004,Tosner2009}; auxiliaries including determination of spin systems (\textsc{simmol}) \cite{Bak2002}; efficient powder averaging \cite{Bak1997,Hohwy1999}; optimisation of experiments for studies of membrane proteins in oriented lipid bilayers \cite{Vosegaard2002}; integration with other software such as \textsc{e}asy\textsc{nmr} \cite{Juhl2020, Koppe2025} for easier fitting of experimental data. A number of reviews and special aspects have also been published \cite{Nielsen2010,Vosegaard2010a,Tosner2017b,Vosegaard2010b}.

\textsc{Epr} is primarily concerned with electron spin dynamics and is theoretically similar to \textsc{nmr}, although electron dynamics occur on different time-scales to nuclear dynamics due to significantly stronger electron spin interactions. An interesting hybrid of \textsc{epr} and \textsc{nmr} is the emerging field of \textsc{dnp}, which utilises the benefits of \textsc{epr} to enhance information content and sensitivity of \textsc{nmr}. This extension of \textsc{simpson} incorporates electron spin dynamics features to enable simulation of pulsed \textsc{epr} and \textsc{dnp} experiments; however, in its present form, it does not include all aspects of \textsc{epr} found in more extensive simulation packages such as EasySpin \cite{Stoll2006,Stoll2014}. It should be noted that many modalities of quantum technologies, in particular quantum sensing, exploit unpaired electron (or exciton) and nuclear spins; this implies that software enabling \textsc{nmr}, \textsc{epr}, and \textsc{dnp} simulations, as proposed here, may find important applications outside these specific disciplines.

The demands dictated by the simultaneous needs for efficiency, accuracy, and versatility in simulation have increased as experiments have become more sophisticated and the effectiveness of \textsc{nmr}, \textsc{epr}, and \textsc{dnp} hardware has improved. The most recent major release of \textsc{simpson}, \textit{version 4.0} \cite{Tosner2014}, focussed on mitigating these numerical demands and was highly efficient in spin dynamics calculations. However, that was more than a decade ago; an interval equivalent to a century in the current computing age. We note that an intermediate version, \textsc{simpson} \textit{version 5.0}, was developed to enable optimal control design of experiments optimising for time-dependent radio-frequency inhomogeneity \cite{Tosner2017a,Tosner2018}. In addition to enabling numerical simulation of \textsc{epr} and \textsc{dnp} pulse sequences, this new major release of \textsc{simpson}, \textit{version 6.0}, includes an improvement in efficiency of numerical calculation for spin dynamics based in the area of propagator splitting \cite{McLachlan2002} to perform rotations and time-propagation operations, 
directly addressing the computational overhead inherent in simulating complex multi-spin ensembles. A primary bottleneck in magnetic resonance simulations arises during the time-evolution of the density matrix, particularly within iterative procedures like optimal control. Such tasks necessitate the repeated evaluation of matrix exponentials, a process that becomes prohibitively expensive as the dimensionality of the Hilbert space increases. Using propagator splitting, it is possible to significantly mitigate this bottleneck and achieve faster simulations without compromising numerical accuracy. Furthermore, new aspects of optimal control, calculation of pulse transients, and higher-order analysis of quadrupolar spin systems are addressed in this release.

\section{SIMPSON: Core aspects}

In a change to previous versions of \textsc{simpson}, which were written in the C programming language, this version of \textsc{simpson} is written in \CC{} and includes many \textit{object-oriented} principles (\textit{vide infra}). This is intended to make the programme more efficient, calculations faster, and the software more compatible with auxiliary components. Furthermore, through the \CC{} architecture this makes programming \textsc{simpson} more user-friendly with the hope for an important increase in community-driven development of \textsc{simpson}; an attractive alternative to the widespread use of in-house software that has a much higher threshold for public dissemination.

\subsection{Using and programming SIMPSON}

Further to the change in programming language, \textsc{simpson} is now released as source code at Gitlab (\url{https://gitlab.au.dk/nmr/simpson}) to be compiled by the user. This is in addition to the programme remaining available as a compiled code for Linux, Windows, and Mac platforms at \url{https://nmr.au.dk}. To further facilitate easy access, \textsc{simpson} is also released via the Docker container platform \cite{docker} (available and ready to use based on the Linux binary, Docker image: {\UrlFont vosegaard/simpson}). For more information about how to use the Docker version see our web site \url{https://nmr.au.dk}). \textsc{simpson} is also available on \textsc{nmr}box \cite{Maciejewski2017} \url{https://nmrbox.nmrhub.org}, enabling direct computation without downloading binaries and providing access to grid-based computing resources. 

\textsc{Tcl} (tool command language) \cite{Tcl} remains the scripting language for \textsc{simpson}. The motivation for maintaining \textsc{tcl} as the controlling scripting language resides in: ($i$) the desire to support the extensive user base that has developed \textsc{simpson} input files over the past quarter of a century, an integral part of pulse sequence development and data extraction, without imposing a significant barrier to the adoption of another language; ($ii$) the technical challenge, far beyond \textsc{tcl}'s utility as a front-end script, arising from the tight coupling between \textsc{tcl} and the remaining C-based components of the core (not \CC{}) to ensure efficient scripting-based interaction with fast compiled software; and ($iii$) the proficient threading capability of \textsc{tcl} for parallelization, which offers advantages over scripting languages such as Python and acts as a key component in achieving fast executable calculations. It is acknowledged that this may present barriers for new users trained in other scripting languages; however, the basic syntax across languages is sufficiently similar to allow for easy translation (aided, for example, by large language model (\textsc{llm}) assistance), supplemented by the numerous \textsc{simpson} input file examples in the literature.

As indicated above, an important component of this \textsc{simpson}\textit{-v6.0} package is the object-oriented principles intended to strengthen the possibilities for community-driven development. Significant effort has been dedicated to overloading operators and functions, so the scientific user need not be concerned with the `deep, dark, rabbit-hole' of linear algebra, the `perpetual Hatter's tea-party' of pointers and memory management, or the complexities of navigating multi-threading to avoid the situation where ``if you don't know where you are going, any road can take you there''. These object-oriented principles may, for example, benefit the \textit{solo user} who should find it much easier to programme novel pulse sequences and develop tools within the \textsc{simpson} framework, rather than developing stand-alone software. Details, instructions, and the potential `pool of tears' involved in compiling \textsc{simpson} source-code will be appended to the software itself, as it is a technical procedure with numerous considerations involving the operating system and the availability of third-party libraries.

\subsection{Visualisation and integration with other software}

\textsc{Simpson}\textit{-v6.0}, like the original version of \textsc{simpson}, is accompanied by the \textsc{simplot} application \cite{Bak2000}, serving as a simple graphical viewer for the display and manipulation of one or more $1$D free-induction decays (\textsc{fid}s), spectra, acquisition data, or output from parameter scans. \textsc{Simplot} is currently only available as binaries at \url{https://nmr.au.dk}, in \textsc{nmr}box \cite{Maciejewski2017}, and in Docker \cite{docker} \textsc{simpson} packages; this is due to technical difficulties associated with the compatibility of the underlying Tk libraries. Accordingly, two modern alternatives are recommended: the \textsc{e}asy\textsc{nmr} platform \cite{Juhl2020, Coulon2025} and a Python-based SimView application.

\begin{figure*}[ht!]
\centering{\includegraphics{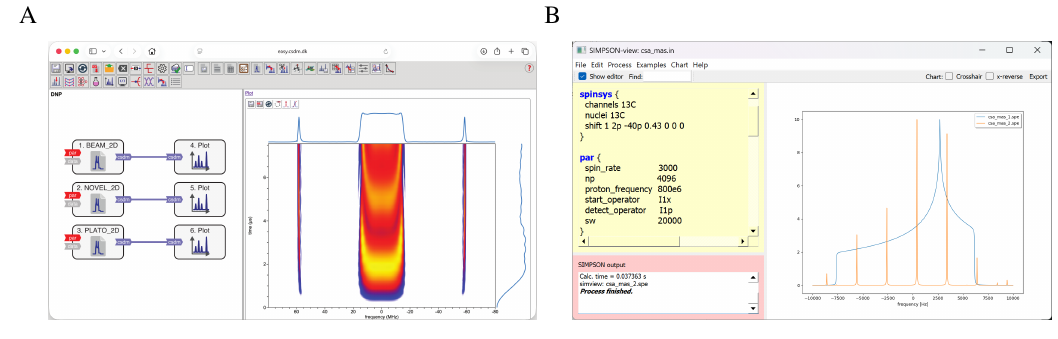}}
\caption{Examples of visualisation of and data integration with \textsc{simpson} simulations. (A) The \textsc{e}asy\textsc{nmr} workflow with three different \textsc{simpson} simulations. The left frame shows the three simulation objects, each forwarding data to a plotting object. The right frame shows the content of each object, here highlighting the plot of one of the simulations. (B) An example of SimView, showing the layout of the interface: a text editor where the \textsc{simpson} input file can be modified; a text output from a running \textsc{simpson} calculation; and a graphical area where calculated \textsc{fid}s or spectra can be plotted.}
\label{FIG_easyNMR_simview}
\end{figure*}

\textsc{E}asy\textsc{nmr} \cite{Juhl2020, Coulon2025} is an online web-based system used to control the workflows involved in visualisation and analysis of general \textsc{nmr} data. It provides a flow-based work environment where the flow objects are connected through \textit{anchors}; the principle is that each object has a specific function and propagates predefined data to the subsequent object. For example, a spectrum is channelled to a modelling object and then onwards to a plotting object. \textsc{E}asy\textsc{nmr} focusses on seamless and lossless transfer of scientific data, relying on the core scientific data model \cite{Srivastava2020}. It is currently hosted on a popular \textsc{nmr}box server \cite{Maciejewski2017}, ensuring a long-term availability and performance. Many features of \textsc{e}asy\textsc{nmr} (e.g. basic spectrum modelling) are performed within the browser, providing flexibility and speed \cite{Koppe2025, Coulon2025}, while other features, such as \textsc{simpson} simulations, are executed seamlessly through the back-end server; this eliminates the need for a local \textsc{simpson} installation \cite{Afrough2025}. The use of \textsc{e}asy\textsc{nmr} to facilitate the running of \textsc{simpson} simulations is described in Ref.~\cite{Juhl2020}. An example of an \textsc{e}asy\textsc{nmr} workflow is illustrated in Fig.~\ref{FIG_easyNMR_simview}A; the left-hand part of the graphical user interface (\textsc{gui}) represents \textsc{beam}, \textsc{novel}, and \textsc{plato} \textsc{dnp} pulse sequence input files exploring excitation pulsing time (vertical) and offset dependency (horizontal), with the \textsc{beam} profile shown to the right (\textit{vide infra}).

SimView (or \textsc{simpson}-view) is a lightweight application, written in Python, serving as a simple front-end for \textsc{simpson} calculations. It is intended to allow the simulations to be executed from a graphical environment as an alternative to the command line utilised by the original \textsc{simpson} and \textsc{simplot} software. The \textsc{gui} comprises three areas: ($i$) a simple text editor where the \textsc{simpson} input file can be modified; ($ii$) an area reserved for console text output from a running \textsc{simpson} calculation; and ($iii$) a graphical area where calculated \textsc{fid}s or spectra can be visualised through simple plots. When properly configured, SimView executes \textsc{simpson} on the active input file as an external process. The resulting \textsc{fid}s or spectra are recognised by SimView with the aid of the text output from \textsc{simpson} using the \simpson{simview} keyword. The data are plotted automatically in the graphical area, which offers interactive manipulation such as zooming, scaling, and the use of a cross-hair cursor, comparison of multiple lines etc., and also facilitates image file generation (using the Python Matplotlib package in a way similar to ssNake \cite{vanMeerten2019}). An example of the SimView layout is presented in Fig.~\ref{FIG_easyNMR_simview}B. The package is available for download from \url{https://github.com/zdetos/Simpson-View}; its open-source code may serve as a basis for embedding \textsc{simpson} calculations within Jupyter notebooks.

Both \textsc{e}asy\textsc{nmr} and SimView are provided as useful tools, particularly for teaching purposes; however, they are accompanied by a warning that they are not intended to exploit or visualise all of the intricate features that \textsc{simpson} can offer. It should be noted that \textsc{e}asy\textsc{nmr} also offers a convenient method for fitting experimental data to \textsc{simpson} simulated data, aiming to extract \textsc{nmr} interaction parameters relating to molecular structure. Alternatively, fitting may be accomplished within a standard \textsc{simpson} script using the \textsc{opt}-1 optimisation package, as described in Ref.~\cite{Tosner2014} and available at \url{https://nmr.au.dk}. 

It should be noted that the \textsc{simmol} software \cite{Bak2002}, which enables the establishment of anisotropic nuclear spin tensors from protein structures (typically \textsc{pdb} files \cite{Berman2000}), is also available via Docker \cite{docker} through the image {\UrlFont vosegaard/simpson}. Alternatively,  \textsc{e}asy\textsc{nmr} has extended capabilities to generate spin systems from various molecular files (e.g. \textsc{pdb}, \textsc{cif}, and \textsc{mol})..

\subsection{Syntax}

There are two interpretations of \textit{syntax} when discussing \textsc{simpson}; the first is the \textsc{tcl}-level syntax that a user programmes as input files, the second is the \CC{} syntax coded within the deep layers of \textsc{simpson} as overloads, routines, functions, and classes. The syntax presented here is only concerned with the former. There are few changes to \textsc{simpson} syntax at the \textsc{tcl}-level; these are primarily additions rather than replacements, intended to encapsulate the inclusion of electron spins for \textsc{epr} and \textsc{dnp} calculations. This syntax is readily understood by current users of \textsc{simpson}, as it follows the same format as that utilised for nuclei. Other new syntax facilitates fast calculations involving propagator splitting, pulse transients, and new optimal control features. A summary of new and modified syntax is outlined in Table~\ref{tab_commands}, and the following sections will introduce these elements through representative examples. For older syntax elements, reference should be made to previous papers on \textsc{simpson} \cite{Bak2000,Tosner2009,Tosner2014}. We note that in this paper, notations and definitions (theory, variables, units, commands etc.) deliberately follow those presented in previous works as closely as possible to facilitate a smooth transition into the new features.

\begin{table*}[!p]\caption{Summary of new features and commands in \textsc{simpson}.}\label{tab_commands}
\renewcommand{\arraystretch}{0}
\setlength{\tabcolsep}{0pt}
\begin{tabular}{l l} 
\hline
\rowcolor{myLightGrey} Definitions in the \simpson{spinsys} section: & \tabdesc{} \\
\rowcolor{myLightGrey} \simpson{channels} $N_1$ $\!N_2$ $\dots$ $N_n$ & \tabdesc{A MW channel of the experiment, for a spin-$\frac{1}{2}$ electron, is introduced as \simpson{e}.} \\
\rowcolor{myLightGrey} \simpson{nuclei} $N_1$ $\!N_2$ $\dots$ $N_n$ & \tabdesc{An \simpson{e} is introduced to declare a spin-$\frac{1}{2}$ electron relevant to the spin system (it is noted that \simpson{nuclei} is a misnomer).} \\
\rowcolor{myLightGrey} \simpson{gtensor} $N$ $\omega_\text{iso}^{\text{eZ}}/\!(2\pi)$ $\omega_\text{aniso}^{\text{eZ}}/\!(2\pi)$ $\eta$ $\alpha$ $\beta$ $\gamma$ & \tabdesc{Defines the electron $g$-tensor interaction (hertz), similar to shift for nuclear spins. $N$ must be an electron.} \\
\rowcolor{myLightGrey} \simpson{hyperfine} $N_1$ $\!N_2$ $a_\text{iso}/\!(2\pi)$ $b_\mathrm{IS}/\!(2\pi)$ $\alpha$ $\beta$ $\gamma$ & \tabdesc{Defines the electron-nuclear hyperfine interaction (hertz). $N_1$ must be an electron, $N_2$ must be a nuclear spin.} \\
\rowcolor{myLightGrey} \simpson{hyperfine_ave} $N_1$ $\!N_2$ $a_\text{iso}/\!(2\pi)$ $b_\mathrm{IS}/\!(2\pi)$ $\eta$ $\alpha$ $\beta$ $\gamma\,\,$ & \tabdesc{The same as \simpson{hyperfine} but allows to define an asymmetry, $\eta$.} \\
\rowcolor{myLightGrey} \simpson{heisenberg} $N_1$ $\!N_2$ $J_\mathrm{ex}$ & \tabdesc{Heisenberg, electron-electron spin exchange, interaction (hertz). Similar to nuclear spin isotropic $J$-coupling.} \\
\rowcolor{myLightGrey} \simpson{edipole} $N_1$ $\!N_2$ $b_\mathrm{SS}/\!(2\pi)$ $\alpha$ $\beta$ $\gamma$ & \tabdesc{Dipole-dipole interaction between two electrons (hertz).} \\
\rowcolor{myLightGrey} \simpson{edipole_ave} $N_1$ $\!N_2$ $b_\mathrm{SS}/\!(2\pi)$ $\eta$ $\alpha$ $\beta$ $\gamma$ & \tabdesc{The same as \simpson{edipole} but allows to define an asymmetry, $\eta$.} \\
\hline
\rowcolor{myLightGrey} Parameters for the \simpson{par} section: & \tabdesc{} \\
\rowcolor{myLightGrey} \simpson{magnetic_field} $B_0$ & \tabdesc{Magnetic field strength, or magnetic induction (tesla), to determine Larmor frequencies. It is equivalent to \simpson{proton_frequency}. If both are given, \simpson{magnetic_field} is used.} \\
\rowcolor{myLightGrey} \simpson{oc_optim_method} value & \tabdesc{Optimisation method selection, alternative to the old \simpson{oc_method} parameter. Can be \simpson{CG} (default), \simpson{L-BFGS}, or \simpson{SIMPLEX}. If the legacy syntax is supplied, it will be used, unless both are supplied, then the new syntax will take precedence. The \simpson{SIMPLEX} method is a gradient-free optimisation method and can be used in the same way as the other optimisation method, but does not require a \simpson{proc gradient} function.} \\
\rowcolor{myLightGrey} \simpson{oc_silent} $N$ & \tabdesc{Silences internal \textsc{simpson} terminal outputs from the optimisation (integer: $0$ or $1$).} \\
\rowcolor{myLightGrey} \simpson{rfmap} filename & \tabdesc{Triggers calculation assuming spatial distribution of RF/MW fields including \textsc{mas} induced temporal modulations. Must be used together with \simpson{pulse_shaped_rotormodulated}. The actual distribution data are loaded from the given file. Time discretisation of the modulations must be equal to or coarser than the time resolution of shaped pulses (\simpson{spin_rate} scales the time axis for modulations).} \\
\rowcolor{myLightGrey} \simpson{method} value1 value2 ... & \tabdesc{Values \simpson{ROTframe}, \simpson{DNPframe}, and \simpson{LABframe} are introduced to declare whether the calculation is conducted in the nuclear spin rotating frame, a \textsc{dnp} frame with electrons in the electron spin rotating frame and nuclear spins in the laboratory-frame, and a laboratory-frame for nuclear spins, respectively. Different form of nuclear interactions is assumed accordingly, electron interactions are available only in its rotating frame form. When \simpson{prop_split} is given, the splitting method is used for time propagation.} \\
\rowcolor{myLightGrey} \simpson{split_order} $N$ & \tabdesc{Defines the propagator splitting order (accuracy) when \simpson{method prop_split} is requested (integer: $0$, $1$, $2$, $3$, $4$, $5$, or $6$).} \\ 
\hline
\rowcolor{myLightGrey} Commands for the \simpson{pulseq} section: & \tabdesc{} \\
\rowcolor{myLightGrey} \simpson{pulse_shaped_rotormodulated} $\delta t$ \textit{sh1} \textit{sh2} \ldots & \tabdesc{Applies shaped pulses on the respective RF channels assuming a spatial RF field distribution including \textsc{mas} induced temporal modulations (\simpson{rfmap} parameter must be properly set). All RF shapes are modified (amplitude \& phase) according to \simpson{rfmap} data.} \\
\hline
\rowcolor{myLightGrey} Commands for the \simpson{main} section: & \tabdesc{} \\ 
\rowcolor{myLightGrey} $\varphi_{nm\!}^{}$ \simpson{create_distortion_operator} file $\delta t$ $\!N$ $\!\Delta t\,\,$ & \tabdesc{Creates an internal representation of distortion operator $\varphi_{nm}$ from impulse response function digitised with time step $\delta t$ and stored in file, applicable on shapes with $N$ pulse elements with element duration $\Delta t$. Time intervals must fulfil $\Delta t = m \delta t$.} \\
\rowcolor{myLightGrey} \simpson{free_distortion_operator} $\varphi_{nm}$ & \tabdesc{Free the internal memory allocated to distortion operator $\varphi_{nm}$.} \\
\rowcolor{myLightGrey} \simpson{distort_shape} $\varphi_{nm}$ $sh^{(p)}$ $sh^{(q)}$ & \tabdesc{Applies a previously defined distortion operator $\varphi_{nm}$ on shape $sh^{(p)}$ to create distorted and finely digitised shape $sh^{(q)}$. The shape $sh^{(q)}$ must be allocated previously.} \\
\rowcolor{myLightGrey} \simpson{reconstruct_gradient} $g^{(q)}$ $g^{(p)}$ $\varphi_{nm}^{1}$ $\varphi_{nm}^{2}$ \ldots & \tabdesc{To be used in the \simpson{gradient} section mainly. Transforms optimal control gradient $g^{(q)}$ calculated with respect to pulse parameters of distorted shapes $sh^{(q)}$ to optimal control gradient $g^{(p)}$ with respect to original pulse parameters of shapes $sh^{(p)}$ that are to be optimised. This transformation is carried out using distortion operators  $\varphi_{nm}^{i}$ used previously to generate the distorted shapes.} \\
\rowcolor{myLightGrey} \textit{value} \simpson{oc_optimize_phase} \textit{sh1} \textit{sh2} \ldots & \tabdesc{Performs phase-only \textsc{grape} optimisation on shaped pulses \textit{sh1} \textit{sh2} \ldots, returning the value of target function and updated shaped pulses \textit{sh1} \textit{sh2} \ldots.} \\
\rowcolor{myLightGrey} \simpson{simview} file1 file2 ... & \tabdesc{To be used within SimView \textsc{gui} application. It triggers display of simulation results, either \textsc{fid}s or spectra, which were generated during \textsc{simpson} calculation and saved on the hard-drive as file1, file2,...} \\
\rowcolor{myLightGrey} \textit{value} \simpson{gval2gtensor} $g_x$ $g_y$ $g_z$ $B_0$ & \tabdesc{For a given set of $g$-values and magnetic field induction $B_0$ (tesla), it calculates \simpson{gtensor} parameters $\omega_\text{iso}^{\text{eZ}}/(2\pi)$, $\omega_\text{aniso}^{\text{eZ}}/(2\pi)$, $\eta$ and stores them as a list in \textit{value}.} \\
\rowcolor{myLightGrey} \textit{value} \simpson{gtensor2gval} $\omega_\text{iso}^{\text{eZ}}/(2\pi)$, $\omega_\text{aniso}^{\text{eZ}}/(2\pi)$ $\eta$ $B_0$ & \tabdesc{Reciprocal conversion to \simpson{gval2gtensor}. }\\
\rowcolor{myLightGrey} \textit{value} \simpson{hyperfine2hypAB} $a_\text{iso}/(2\pi)$ $b_\mathrm{IS}/(2\pi)$ $\beta_\mathrm{PL}$ & \tabdesc{From a given \simpson{hyperfine} parameters, it calculates secular and pseudo-secular hyperfine constants $A$ and $B$, respectively, stored as a list in \textit{value}.} \\
\hline
\end{tabular}
\end{table*}

Recently, \textsc{simpson}\textit{-v5.0} was utilised as a platform to conduct optimal control calculations involving spatial radio-frequency (RF) field distribution and magic-angle spinning (\textsc{mas}) induced temporal modulations \cite{Tosner2017a}. For this purpose, the parameter \simpson{rfmap} and a new command, \simpson{pulse_shaped_rotormodulated}, were introduced. Their usage is described in the Supplementary Material of Ref.~\cite{Tosner2018} and is included in Table~\ref{tab_commands} for completeness. In addition, a full input file is provided for the optimisation of tm-\textsc{spice} pulses in the Supplementary Material and the supplementary collection of codes, including example data files with spatial RF field distributions typical for \textsc{mas} probes with rotor diameters of $3.2$, $1.3$, and $0.7~\si{\milli\metre}$. These data files can be used as viable values for the \simpson{rfmap} parameter. Importantly, these new features are distinct from the \simpson{rfprof_file} parameter, which characterizes a static distribution of RF field values \cite{Tosner2009}.

\section{Electron spin interactions}

While nuclear spin interactions have been described in detail in the original \textsc{simpson} paper \cite{Bak2000} and subsequent accounts, the electron spin interactions necessary for the calculation of \textsc{epr} and \textsc{dnp} experiments require formal definition. For consistency, a formalism adopted here reflects that utilised for nuclear spin interactions in previous \textsc{simpson} releases \cite{Bak2000,Tosner2009,Tosner2014}. This approach is also consistent with reviews on solid-state \textsc{nmr} dipolar recoupling \cite{Nielsen2012,Ladizhansky2024} and is formulated using irreducible tensor operators and Wigner transformations. In this initial description, the scope is restricted to spin-$\frac{1}{2}$ systems. This ensures interactions remain sufficiently weak to remain compatible with the secular approximation within the electron-spin rotating frame.

Following the standard notation of magnetic resonance, the operators of a composite system containing nuclei (labelled with $\mathrm{I}$) and electron (labelled with $\mathrm{S}$) are constructed through Kronecker products with
\begin{equation}
\begin{gathered}
\hat{\mathrm{I}}_{\{\mathrm{x},\mathrm{y},\mathrm{z}\}} \triangleq \mathds{1}\otimes\mathds{1}\otimes\cdots\otimes\frac{1}{2}\hat{\sigma}_{\{\mathrm{x},\mathrm{y},\mathrm{z}\}}\otimes\cdots\otimes\mathds{1}\\
\hat{\mathrm{S}}_{\{\mathrm{x},\mathrm{y},\mathrm{z}\}} \triangleq \mathds{1}\otimes\cdots\otimes\frac{1}{2}\hat{\sigma}_{\{\mathrm{x},\mathrm{y},\mathrm{z}\}}\otimes\cdots\otimes\mathds{1}\otimes\mathds{1}
\end{gathered}\label{EQ_pauli_kron}
\end{equation}
with the identity matrix, $\mathds{1}$, and where the position of the non-identity, the $\hat{\sigma}_{\{\mathrm{x},\mathrm{y},\mathrm{z}\}}$ Pauli matrices, in the Kronecker chain is a unique position for each spin. In what follows, it is assumed that the labelled operators $\hat{\mathrm{S}}$ and $\hat{\mathrm{I}}$, for electrons and nuclei respectively, are related to their corresponding Pauli matrices through an ordered Kronecker product chain.

The electron-spin Zeeman interaction, representing the linear coupling of the electron-spin dipole moment to the external magnetic field, may be described by the Hamiltonian 
\begin{equation}
\mathcal{H}_\mathrm{eZ}^{} = \frac{\beta_\mathrm{e}}{\hbar} \, \hat{\vec{\mathrm{S}}} \cdot \mathbf{g} \cdot \vec{B} \quad ,
 \label{EQ_HZ}
\end{equation}
with $\vec{B}$ denoting the external magnetic field vector, $\hat{\vec{\mathrm{S}}} = \begin{pmatrix} \hat{\mathrm{S}}_\mathrm{x} & \hat{\mathrm{S}}_\mathrm{y} & \hat{\mathrm{S}}_\mathrm{z} \end{pmatrix}$ is a vector of (Cartesian) electron-spin operators, and $\beta_\mathrm{e}$ is related to the Bohr magneton. In its principal-axis-system (\textsc{pas}, henceforth labelled P), the $g$-tensor, $\mathbf{g}$, takes the form
\begin{equation}
\mathbf{g} =
\begin{bmatrix}
g_\mathrm{x} & 0 & 0\\
0 & g_\mathrm{y} & 0\\
0 & 0 & g_\mathrm{z}
\end{bmatrix} \quad .
 \label{EQ_gPAS}
\end{equation}
Using the irreducible spherical tensor formalism, with spatial tensors $\mathcal{R}_{\ell,m}^{}$ and spin tensors $\mathcal{T}_{\ell,m}^{}$, the Hamiltonian in the laboratory-frame (labelled L) can be expressed as 
\begin{gather}
\mathcal{H}_\mathrm{eZ}^{} = \mathcal{C}^{\mathrm{eZ}}\mathcal{R}_{0,0}^{\mathrm{eZ}} \mathcal{T}_{0,0}^{\mathrm{eZ}} + \mathcal{C}^{\mathrm{eZ}} \sum_{m=-2}^{2} (-1)_{}^m (\mathcal{R}_{2,-m}^{\mathrm{eZ}})^{\mathrm{L}} \mathcal{T}_{2,m}^{\mathrm{eZ}}
,\nonumber\\
\mathcal{H}_\mathrm{eZ}^{} \approx \mathcal{C}^{\mathrm{eZ}}\mathcal{R}_{0,0}^{\mathrm{eZ}} \mathcal{T}_{0,0}^{\mathrm{eZ}} + \mathcal{C}^{\mathrm{eZ}} (\mathcal{R}_{2,0}^{\mathrm{eZ}})^{\mathrm{L}} \mathcal{T}_{2,0}^{\mathrm{eZ}} \quad ,
 \label{EQ_Hz_irred}
\end{gather}
with a transformation to the electron-spin Zeeman interaction frame (which is assumed to be along the z-axis) and making the secular (high-field) approximation. The fundamental constant is given by $\mathcal{C}^{\mathrm{eZ}} = \frac{\beta_\mathrm{e}}{\hbar}$. The second-rank spatial tensor in the laboratory-frame relates to the tensor in the principal-axis-frame as
\begin{equation}
(\mathcal{R}_{2,0}^{\mathrm{eZ}})^{\mathrm{L}} = \sum_{m=-2}^{2} \mathcal{D}_{m,0}^{(2)} \Big(\Omega_{}^{\mathrm{PL}}\Big) (\mathcal{R}_{2,m}^{\mathrm{eZ}})^{\mathrm{P}} \quad ,
 \label{EQ_R2transf}
\end{equation}
where $\mathcal{D}_{m,m'}^{(n)}$ denotes the $m,m'$ components of a $n^\text{th}$-rank Wigner matrix with $ \Omega^{\mathrm{PL}}=\big\{\alpha^{\mathrm{PL}},\beta^{\mathrm{PL}},\gamma^{\mathrm{PL}}\big\}$ representing the Euler angles relating the principal-axis to laboratory-frame, of which $\gamma^{\mathrm{PL}}$ is irrelevant in the secular approximation. In presence of several electron-spin interactions, with different principal-axis-frames, it may be relevant to add in a crystal-fixed frame $\mathrm{C}$ (in-between the individual $\mathrm{P}$'s and $\mathrm{L}$) as described earlier in relation to multiple solid-state \textsc{nmr} relevant anisotropic interactions \cite{Bak2000}. The spin tensors are defined as 
\begin{equation}
\mathcal{T}_{0,0}^{\mathrm{eZ}} = B_\mathrm{z}^{}\hat{\mathrm{S}}_\mathrm{z}^{} \quad\text{and}\quad \mathcal{T}_{2,0}^{\mathrm{eZ}} = \sqrt{\frac{2}{3}} B_\mathrm{z}^{}\hat{\mathrm{S}}_\mathrm{z}^{} \quad , \nonumber
\end{equation}
while the spatial tensors are given by 
\begin{gather}
\mathcal{R}_{0,0}^{\mathrm{eZ}} =  g_\text{iso}\quad,\quad (\mathcal{R}_{2,\pm2}^{\mathrm{eZ}})^{\mathrm{P}} = -\frac{1}{2} \eta  \Delta g \quad ,\nonumber\\
(\mathcal{R}_{2,\pm1}^{\mathrm{eZ}})^{\mathrm{P}} = 0,\quad \text{and}\quad (\mathcal{R}_{2,0}^{\mathrm{eZ}})^{\mathrm{P}} = \sqrt{\frac{3}{2}} \Delta g \quad ,\nonumber
\end{gather}
with definitions of the isotropic $g$-value, $g_\text{iso}$, $g$-anisotropy, $\Delta g$, and the asymmetry parameter, $\eta$, as
\begin{equation}
g_\text{iso}=\frac{g_\mathrm{x} + g_\mathrm{y} + g_\mathrm{z}}{3},\quad \Delta g=g_\mathrm{z} - g_\text{iso}, \quad \eta = \frac{g_\mathrm{y} - g_\mathrm{x}}{\Delta g}.\nonumber
\end{equation}
Since \textsc{simpson} calculations are usually performed in the electron-spin rotating frame, it is often more convenient to define the electron Zeeman interaction in frequency units relative to the carrier frequency, analogously to a chemical shift for nuclei. This is why, for the \simpson{gtensor} keyword, the user defines an isotropic shift 
$\omega_\text{iso}^{\mathrm{eZ}}=g_\text{iso}\frac{\beta_\mathrm{e}}{h}B_\mathrm{z}-\omega_\textsc{mw}$, a shift anisotropy $\omega_\text{aniso}^{\mathrm{eZ}}=\Delta g \frac{\beta_\mathrm{e}}{h}B_\mathrm{z}$,  
and the asymmetry $\eta$ instead of the $g$-values themselves. To facilitate conversion between these two representations, conversion routines \simpson{gval2gtensor} and \simpson{gtensor2gval} are provided, with the magnetic field strength (in teslas) as an additional parameter. The Haeberlen convention is used to label and order the principal elements according to $|g_\mathrm{z}-g_\text{iso}| \geq |g_\mathrm{x}-g_\text{iso}| \geq |g_\mathrm{y}-g_\text{iso}|$.

The electron-nuclear spin hyperfine coupling, representing the bilinear interaction between their two magnetic dipoles, is given by the Hamiltonian
\begin{equation}
\mathcal{H}_\mathrm{HF}^{} =  \hat{\vec{\mathrm{S}}} \cdot \mathbf{A} \cdot \hat{\vec{\mathrm{I}}} \quad , 
 \label{EQ_Hhf}
\end{equation}
where $\hat{\vec{\mathrm{I}}}= \begin{pmatrix} \hat{\mathrm{I}}_\mathrm{x} & \hat{\mathrm{I}}_\mathrm{y} & \hat{\mathrm{I}}_\mathrm{z} \end{pmatrix}$ is a vector of (Cartesian) nuclear spin operators. In a single-electron picture, the interaction matrix $\mathbf{A}$ consists of two contributions: the isotropic Fermi-contact and the anisotropic dipole-dipole interactions,
\begin{equation}
 \mathbf{A}  = a_\text{iso} \mathbf{1} + \mathbf{T} \quad , 
 \label{EQ_Ahf}
\end{equation}
with $\mathbf{1}$ being the unit matrix. The Fermi-contact coupling is defined as $a_\text{iso} = \frac{2}{3} \frac{\mu_0}{\hbar} \gamma_\mathrm{e} \gamma_\mathrm{n} |\Psi_0(0)|^2$ where $|\Psi_0(0)|^2$ is the electron spin density at the nucleus point. Within the point-dipole approximation, the electron-nuclear dipolar term $\mathbf{T}$ is represented by a symmetric tensor.
Using irreducible tensor operators and assuming the secular approximation for the electron spin, the hyperfine coupling Hamiltonian in Eq.~(\ref{EQ_Hhf}) may be recast in the spatial laboratory-frame as
\begin{eqnarray}
\mathcal{H}_\mathrm{HF}^{} &\approx& \mathcal{C}^{\mathrm{HF}}_{\text{iso}}\mathcal{R}_{0,0}^{\mathrm{HF}} \mathcal{T}_{0,0}^{\mathrm{HF}} + \mathcal{C}^{\mathrm{HF}}_{\text{dip}} \{ (\mathcal{R}_{2,0}^\mathrm{HF})^\mathrm{L} \mathcal{T}_{2,0}^{\mathrm{HF}} \nonumber \\ &+& (\mathcal{R}_{2,-1}^\mathrm{HF})^\mathrm{L} \mathcal{T}_{2,1}^{\mathrm{HF}} + (\mathcal{R}_{2,1}^\mathrm{HF})^\mathrm{L} \mathcal{T}_{2,-1}^{\mathrm{HF}} \} \label{EQ_Hhfirred_DNP} \quad ,
\end{eqnarray}
using $\mathcal{C}^{\mathrm{HF}}_{\text{iso}} = 1$ and $\mathcal{C}^{\mathrm{HF}}_{\text{dip}} = -2 \gamma_\mathrm{e} \gamma_\mathrm{n} \hbar$, with the gyromagnetic ratios $\gamma_\mathrm{e}$ and $\gamma_\mathrm{n}$ of the electron and nuclear spins, respectively.
The relevant spin operators are truncated to exclude electron-spin transverse components and take the form
\begin{gather}
\mathcal{T}_{0,0}^{\mathrm{HF}} = \hat{\mathrm{S}}_\mathrm{z}^{} \hat{\mathrm{I}}_\mathrm{z}^{}, \quad\mathcal{T}_{2,0}^{\mathrm{HF}} = \sqrt{\frac{2}{3}} \hat{\mathrm{S}}_\mathrm{z}^{} \hat{\mathrm{I}}_\mathrm{z}^{},\quad\mathcal{T}_{2,\pm 1}^{\mathrm{HF}} = \mp \frac{1}{2} \hat{\mathrm{S}}_\mathrm{z}^{} \hat{\mathrm{I}}_\pm^{}.\nonumber
\end{gather}
The laboratory-frame spatial terms again transform as in Eq.~(\ref{EQ_R2transf}), while the corresponding principal-axis-frame spatial terms take the form 
\begin{gather}
\mathcal{R}_{0,0}^{HF} = a_\text{iso},\quad (\mathcal{R}_{2,0}^\mathrm{HF})^\mathrm{P} = \sqrt{\frac{3}{2}} \frac{\mu_0}{4\pi}\frac{1}{r^3} ,\nonumber\\
(\mathcal{R}_{2,\pm1}^\mathrm{HF})^\mathrm{P} = (\mathcal{R}_{2,\pm2}^\mathrm{HF})^\mathrm{P} = 0, \nonumber
\end{gather}
Generally, there are other contributions to the hyperfine coupling between electrons and nuclei that are beyond the scope of the current presentation, resulting in a general form of the coupling matrix $\mathbf{A}$. In that case,
\begin{equation}
a_\text{iso}=\frac{A_\mathrm{xx}\!+\!A_\mathrm{yy}\!+\!A_\mathrm{zz}}{3},\;\; \Delta a=A_\mathrm{zz}-a_\text{iso},\;\; \eta=\frac{A_\mathrm{yy}\!-\!A_\mathrm{xx}}{\Delta a}\nonumber
\end{equation}
are defined with corresponding principal-axis-frame spatial terms $(\mathcal{R}_{2,m}^\mathrm{HF})^\mathrm{P}$ in the same form as for the $g$-tensor, while substituting $a_\text{iso}$ and $\Delta a$ for $g_\text{iso}$ and $\Delta g$, respectively, and assuming $\mathcal{C}^{\mathrm{HF}}$ constants equal to unity.

To define the hyperfine interaction in \textsc{simpson}, the keyword \simpson{hyperfine} is used; this combines the Fermi-contact and dipolar contributions through values $a_\text{iso}$ and $b_\mathrm{IS}^{}$, where the latter is given as $b_\mathrm{IS}^{}=-\frac{\mu_0}{4\pi}\frac{\hbar \gamma_\mathrm{e}\gamma_\mathrm{n}}{r^3}$. In the more general case of a possibly non-zero asymmetry, the corresponding keyword is \simpson{hyperfine_ave}, which adopts the values  $a_\text{iso}$, $b_\mathrm{IS} = \Delta a/2$, and $\eta$ (here, the $b_\mathrm{IS}$ constant should be understood as an effective dipolar constant). 

The truncated hyperfine Hamiltonian is frequently represented as 
\begin{equation}
\mathcal{H}_\mathrm{HF}^{} = A \hat{\mathrm{S}}_\mathrm{z}^{} \hat{\mathrm{I}}_\mathrm{z}^{} + B \hat{\mathrm{S}}_\mathrm{z}^{} \hat{\mathrm{I}}_\mathrm{x}^{} \quad ,
\end{equation}
employing secular ($A$) and pseudo-secular ($B$) hyperfine coupling constants related to the hyperfine coupling tensor defined above as $A=A_\mathrm{zz}=a_\text{iso}+b_\mathrm{IS}(3\cos^2{\beta_\mathrm{PL}}-1)$ and $B=\frac{3}{2}b_\mathrm{IS}\sin{2\beta_\mathrm{PL}}$, where $\beta_\mathrm{PL}$ is the angle between the static magnetic field and the symmetry axis of the hyperfine coupling tensor. These values are calculated using the conversion routine \simpson{hyperfine2hypAB} included in this \textsc{simpson}\textit{-v6.0} release.

The Hamiltonian for the direct dipole-dipole interaction between two electron spins has the same form as for the homonuclear dipole-dipole interactions
\begin{equation}
\mathcal{H}_\mathrm{eD}^{} \approx \mathcal{C}^{\mathrm{eD}}  (\mathcal{R}_{2,0}^{\mathrm{eD}})^{\mathrm{L}} \mathcal{T}_{2,0}^{\mathrm{eD}} \quad ,
 \label{EQ_HeDeD_irred}
\end{equation}
with $\mathcal{C}^{\mathrm{eD}}=-2\hbar \gamma_\mathrm{e}^2$, the spin operator 
\begin{equation}
\mathcal{T}_{2,0}^{\mathrm{eD}} = \sqrt{\frac{1}{6}} \Big( 2\hat{\mathrm{S}}_\mathrm{1z}^{} \hat{\mathrm{S}}_\mathrm{2z}^{}-\hat{\mathrm{S}}_\mathrm{1x}^{}\hat{\mathrm{S}}_\mathrm{2x}^{}-\hat{\mathrm{S}}_\mathrm{1y}^{}\hat{\mathrm{S}}_\mathrm{2y}^{}\Big)\quad ,\nonumber
\end{equation}
and the principal-axis-frame spatial operators
\begin{gather}
(\mathcal{R}_{2,0}^{\mathrm{eD}})^{\mathrm{P}} = \sqrt{\frac{3}{2}} \frac{\mu_0}{4\pi}\frac{1}{r^3} ,\quad \mathcal{R}_{2,\pm1}^{\mathrm{P}} = (\mathcal{R}_{2,\pm2}^{\mathrm{eD}})^{\mathrm{P}} = 0.\nonumber
\end{gather}
The subscripts, $k\in\{1,2\}$, of the operators $\hat{\mathrm{S}}_\mathrm{k\{\mathrm{x},\mathrm{y},\mathrm{z}\}}$ label each of the two electron spins. The Hamiltonian in Eq.~(\ref{EQ_HeDeD_irred}) is defined using the keyword \simpson{edipole} and the dipolar coupling constant $b_\mathrm{SS}^{}=-\frac{\mu_0}{4\pi} \frac{\hbar \gamma_\mathrm{e}^2}{r^3}$. In the special case of non-zero asymmetry, the variant \simpson{edipole_ave} is available in Table~\ref{tab_commands}; this is implemented in the same way as the nuclear \simpson{dipole_ave} to enable $\eta$ to be defined.

Finally, the Heisenberg coupling (exchange) interaction between two electron spins, which resembles the isotropic homonuclear $J$-coupling known from \textsc{nmr}, is described by the Hamiltonian
\begin{equation}
\mathcal{H}_\mathrm{ex}^{} = J_\mathrm{ex}^{}  \hat{\vec{\mathrm{S}}}_1^{} \cdot \hat{\vec{\mathrm{S}}}_2^{}  = C^{\mathrm{ex}}\mathcal{R}_{0,0}^{\mathrm{ex}} \mathcal{T}_{0,0}^{\mathrm{ex}} \quad ,
 \label{EQ_Hex}
\end{equation}
where $J_\mathrm{ex} = \mathcal{R}_{0,0}^{\mathrm{ex}}$ denotes the isotropic exchange coupling constant, and $C^{\mathrm{ex}}$ is equal to unity to conform to the notation. The Heisenberg interaction may, for example, be significant if radicals are sufficiently close for wave-function overlap. No further simplification or truncation is assumed when the two electron spins have different $g$ values and under high-field conditions. The interaction in Eq.~(\ref{EQ_Hex}) is defined using the \simpson{heisenberg} keyword; this was named to avoid confusion with chemical exchange, which may be relevant for future implementations in \textsc{simpson}.

As indicated by the notation used above, \textsc{Simpson} performs the simulation of \textsc{epr} and \textsc{dnp} experiments with electron spins in their rotating frame and with any nuclear spins in the laboratory frame; the option \simpson{DNPframe} \textit{must} be defined in \simpson{par(method)} for \textsc{dnp} and \textsc{epr} simulations.

\subsection{Electron paramagnetic resonance}

\begin{figure}[ht!]
\centering{\includegraphics{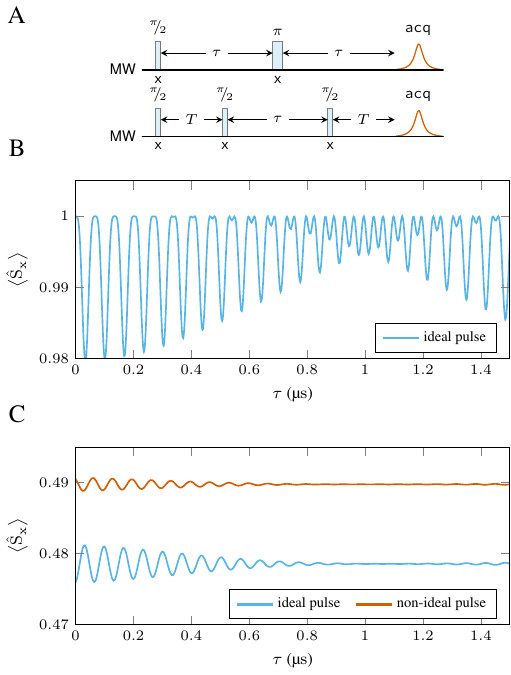}}
\caption{(A) Pulse sequence diagrams for 2-pulse and 3-pulse \textsc{eseem} experiments. (B) Simulated echo intensity, $\langle \hat{\mathrm{S}}_x \rangle$), as a function of the delay time $\tau$ for 2-pulse \textsc{eseem} for a single orientation of an electron-proton two-spin system with ideal pulses (using \simpson{pulseid}). (C) Corresponding simulation for 3-pulse \textsc{eseem} using the same spin system, but with a summation over several crystallites. Additionally, an explicit summation over electron offsets is performed. The effect of using non-ideal (soft) pulses (red, using \simpson{pulse}) is shown relative to ideal-pulses (blue, using \simpson{pulseid}).}
\label{FIG_eseem}
\end{figure}

To illustrate the use of \textsc{simpson} for the simulation of pulsed \textsc{epr} experiments, Fig.~\ref{FIG_eseem} provides examples of electron echo envelope modulation (\textsc{eseem}) \cite{Rowan1965}. These include representative demonstrations of single-crystal orientations, powder averaging, and offset averaging, evaluated in the context of both ideal pulses and finite-duration (physically real) pulses.

Starting with a simple electron-nuclear two-spin system, the \textsc{simpson} syntax proceeds to create a \simpson{spinsys} with relevant interactions using the following syntax:
\begin{lstlisting}[style=SIMPSON-coloured]
spinsys {
    channels    e
    nuclei      e 1H
    gtensor     1 0 0 0 0 0 0
    hyperfine   1 2 0 1e6 0 45 0
}
\end{lstlisting}
This specification, as detailed in Table~\ref{tab_commands}, utilises the misnomer of categorizing an electron spin within a \simpson{nuclei} entry. This convention is maintained to ensure the framework remains conceptually consistent with \textsc{nmr} simulations. The example system comprises an electron spin and a $^{1}\text{H}$ nucleus, whereby irradiation is defined on the electron-spin channel. In this particular case, the $g$-tensor (expressed in $\si{\hertz}$) is isotropic and configured to match the rotating-frame frequency exactly. The hyperfine interaction has no isotropic component, an anisotropy of $1~\si{\mega\hertz}$, and the Euler angle $\beta_\mathrm{PL}$ of $45\si{\degree}$. The Hamiltonian for this system is formulated as:
\begin{equation}
\mathcal{H}(t)=\omega_\mathrm{x}^{}(t)\hat{\mathrm{S}}_{\mathrm{x}}^{}+\omega_\mathrm{y}^{}(t)\hat{\mathrm{S}}_{\mathrm{y}}^{} + \Delta \omega_\mathrm{S}^{} \hat{\mathrm{S}}_{\mathrm{z}}^{} + A \hat{\mathrm{S}}_{\mathrm{z}}^{} \hat{\mathrm{I}}_{\mathrm{z}}^{} + B \hat{\mathrm{S}}_{\mathrm{z}}^{} \hat{\mathrm{I}}_{\mathrm{x}}^{} \quad ,
\label{EQ_spinsys1}
\end{equation}
where all frequencies are expressed in angular units. The time-dependent $x-$ and $y-$phase MW irradiation and electron spin offset are denoted by the symbol $\omega$; this characterizes the angular frequency and ensures consistency with previous \textsc{simpson} accounts \cite{Bak2000,Tosner2009,Tosner2014}. For instance, the internal electron spin interactions are categorized according to the widely utilised secular and pseudo-secular forms.
 
The hyperfine coupling tensor parameters defined in \simpson{spinsys} can be converted to the secular and pseudo-secular hyperfine constants $A$ and $B$, respectively, using the procedure \simpson{hyperfine2hypAB}. In this specific example, the conversion routine:
\begin{lstlisting}[style=SIMPSON-coloured]
    set valAB [hyperfine2hypAB 0 1e6 45]
    puts "A = [lindex $valAB 0] Hz"
    puts "B = [lindex $valAB 1] Hz"
\end{lstlisting}
outputs the values $A=0.5~\si{\mega\hertz}$ and $B=1.5~\si{\mega\hertz}$ (following insertion in a \textsc{tcl} script, e.g. \simpson{ABconv.in}, and execution in a shell using \textsc{simpson} in the same manner as standard input files).  
 
For the given examples, the pulses $\omega_\mathrm{x,y}^{}(t)$ in Eq.~(\ref{EQ_spinsys1}) are time-dependent MW amplitudes as defined by the pulse sequences in Fig.~\ref{FIG_eseem}A, applied to the electron spin with an offset $\Delta \omega_\mathrm{S}^{}$ = $\omega_\mathrm{S}^{}-\omega_\textsc{mw}^\text{carrier}$ relative to the MW carrier frequency (the latter takes the value $0$ in the \simpson{spinsys} above, but is included for further examples that may be off-resonance).

The common parameters for simulating the pulsed \textsc{eseem} experiments shown in Fig.~\ref{FIG_eseem}A are set in the \simpson{par} section of \textsc{simpson} code:
\begin{lstlisting}[style=SIMPSON-coloured]
par {
    proton_frequency    14.8e6
    start_operator      I1x
    detect_operator     I1x
    method              DNPframe
    np                  384
    sw                  250
}
\end{lstlisting}
These parameters assume X-band conditions ($9.742~\si{\giga\hertz}$ for the electrons) corresponding to $14.8~\si{\mega\hertz}$ Larmor frequency for protons. An additional option, \simpson{DNPframe}, has been added to \simpson{method} and must be defined for \textsc{dnp} and \textsc{epr} simulations (instead of \simpson{ROTframe} or \simpson{LABframe}). The $\frac{\pi}{2}$-pulse is not explicitly calculated, with the starting and detection operators set along the $x$-axis. The variables \simpson{np} and \simpson{sw} (the number of acquisition points and the spectral width, respectively) are internal \textsc{simpson} parameters; in this instance, \simpson{sw} only influences the simulation when writing values to the output file. In the absence of a \simpson{crystal_file}, a single-crystal orientation is determined using the provided $\mathrm{P}$ to $\mathrm{L}$ Euler angles.

For the 2-pulse \textsc{eseem} experiment in Fig.~\ref{FIG_eseem}A (top row), the \simpson{pulseq} section is written in \textsc{simpson} as:
\begin{lstlisting}[style=SIMPSON-coloured]
proc pulseq {} {
    global par
    # set delay incr., max. delay, then loop over par(np)
    set dt 0.004
    set T [expr $par(np)*$dt]
    for {set tau 0} {$tau<$T} {set tau [expr $tau+$dt]} {
        reset
        delay   $tau
        pulseid 0.005 100e6 x
        delay   $tau
        acq
    }
}
\end{lstlisting}
where the loop over \simpson{delay} values is performed explicitly in \textsc{tcl}. The pulse sequence (initialised to \simpson{start_operator I1x}) consists of a \simpson{delay} of \simpson{tau}, followed by and ideal pulse, \simpson{pulseid}, and a further \simpson{delay} of \simpson{tau}. The flip angle of the pulse is calculated using $5~\si{\nano\second}$ at $100~\si{\mega\hertz}$ around \simpson{x}; \simpson{pulseid} assumes an infinitely short pulse in calculations. If a finite pulse is required, \simpson{pulseid} is replaced with \simpson{pulse}. The \simpson{delay}, \simpson{tau}, is incremented by \simpson{dt} $=4~\si{\nano\second}$ to a maximum of \simpson{T} $=1.53~\si{\micro\second}$. 

The \simpson{main} section is then formulated as:
\begin{lstlisting}[style=SIMPSON-coloured]
proc main {} { fsave [fsimpson] eseem.fid }
\end{lstlisting}
The resulting simulation of an \textsc{eseem} trace, saved in the file \simpson{eseem.fid}, is shown in Fig.~\ref{FIG_eseem}B; in this instance, where only a single orientation is calculated, the pulses are assumed to be ideal.

For 3-pulse \textsc{eseem}, with the pulse sequence given in Fig.~\ref{FIG_eseem}A (lower row), the spin system is identical to that used in the 2-pulse \textsc{eseem} example, with the significant difference being the inclusion of a \simpson{crystal_file} to simulate a powder and an \simpson{averaging_file} to simulate an offset bandwidth. Specifically, the additional syntax lines are appended to the \simpson{par} section:
\begin{lstlisting}[style=SIMPSON-coloured]
    #...
    crystal_file        rep144
    averaging_file      gtensor_1_iso_30MHz.ave
    #...
\end{lstlisting}
Here, the internal \simpson{crystal_file rep144} denotes that the powder is represented by $144$ $\{\alpha_\mathrm{PL}, \beta_\mathrm{PL}\}$ pairs of angles selected using the \textsc{repulsion} method \cite{Bak1997}. The internal \simpson{averaging_file gtensor_1_iso_30MHz.ave} organises the averaging over an offset range specified as a list of offsets and weights in the file \simpson{ gtensor_1_iso_30MHz.ave} (included in the Supplementary Material). In principle, an anisotropic $g$-tensor could also be utilised, but then the anisotropy of the hyperfine interaction is correlated with the anisotropy of the $g$-tensor. It is common in \textsc{epr} to have, for example, unresolved hyperfine couplings to nuclei in the solvent; it is therefore practical to perform simulations over a range of offsets. In this instance, a uniform distribution is utilised (in the specified file), though it would be straightforward to utilise a Gaussian one. Note that for an \simpson{averaging_file} to be used, the parameter being modified must be initialised in the \simpson{spinsys} with a nominal value. In this specific case, \simpson{gtensor 1 1 0 0 0 0 0} (the nominal $\omega_\text{iso}^{\text{eZ}}/(2\pi)$ is set to $1~\si{\hertz}$ but is overwritten at every stage of the simulation with the \simpson{averaging_file}).

The \simpson{pulseq} section for 3-pulse \textsc{eseem}, incorporating an additional pulse and delay relative to 2-pulse \textsc{eseem}, is written as:
\begin{lstlisting}[style=SIMPSON-coloured]
proc pulseq {} {
    global par
    # coherence selection
    matrix set 1 coherence {{0 1} {0 -1} {0 0}}
    # set delay incr., max. delay, then loop over par(np)
    set dt 0.004
    set T [expr $par(np)*$dt]
    for {set tau 0} {$tau<$T} {set tau [expr $tau+$dt]} {
        reset
        delay   0.104
        pulseid 0.0025 100e6 x
        filter  1
        delay   $tau
        pulseid 0.0025 100e6 x
        delay   0.104
        acq
    }
}
\end{lstlisting}
where coherence selection is introduced via the commands:
\begin{lstlisting}[style=SIMPSON-coloured]
    matrix set 1 coherence {{0 +1} {0 -1} {0 0}}
    #...
    filter 1
\end{lstlisting}
These commands select a coherence order of $0$ for the electron spin at the beginning of the free evolution time, while retaining the coherence order $\{+1,0,-1\}$ for the nuclear spin.

The \simpson{main} section identical to that of the 2-pulse \textsc{eseem} example. The resulting simulated signal for the 3-pulse \textsc{eseem} pulse sequence is shown in Fig.~\ref{FIG_eseem}C. Furthermore, the effect of soft pulses is illustrated using \simpson{pulse} (of $25~\si{\nano\second}$ at $10~\si{\mega\hertz}$ around \simpson{x}) instead of \simpson{pulseid} in the \simpson{pulseq} section. As expected, the modulation differs for a non-ideal (soft) pulse whose bandwidth is insufficient to excite both allowed and forbidden transitions in the system \cite{Schweiger2001}.

\subsection{Dynamic Nuclear Polarisation}

Dynamic nuclear polarisation (\textsc{dnp}) \cite{Overhauser1953,Carver1953,Becerra1993} has become a powerful and widely used technique for enhancing nuclear spin polarisation. By transferring the high electron-spin polarisation to the relatively low nuclear-spin polarisation ensemble \cite{ArdenkjaerLarsen2015,LillyThankamony2017,Moroz2022}, experimental sensitivity in \textsc{nmr} can be enhanced by orders of magnitude. This approach enables the simultaneous, indirect probing of electron and nuclear spin information and facilitates applications in quantum sensing. Until recently, \textsc{dnp} followed a similar path to the early development of \textsc{nmr} by utilising \textit{continuous-wave} (CW) irradiation to manipulate the spins; in this instance, MW irradiation is applied to the electron spins as opposed to RF irradiation on the nuclear spins (noting that the term RF also covers most typical MW frequencies utilised in \textsc{epr} and \textsc{dnp}). 

Pulsed realisation of \textsc{dnp} is currently receiving considerable attention, with the potential to yield performance gains comparable to the earlier transition from CW to pulsed \textsc{nmr}. Pulsed \textsc{dnp} experiments for electron-to-nuclear spin polarisation transfer include techniques such as: nuclear-spin orientation via electron-spin locking (\textsc{novel}) \cite{Henstra1988};  integrated solid-effect (\textsc{ise}) \cite{Henstra1988_2}; PulsePol \cite{Schwartz2018}; \textsc{top-dnp} \cite{Tan2019}; X-inverse-X (\textsc{x}i\textsc{x}) \textsc{dnp} \cite{Redrouthu2022}; \textsc{tppm} \cite{Redrouthu2023}; broadband excitation by amplitude modulation (\textsc{beam}) \cite{Wili2022}; polarisation transfer via non-linear optimisation (\textsc{plato}) \cite{Nielsen2024}; constrained random-walk optimisation (c\textsc{rw-opt}) \cite{Nielsen2025}; longitudinal pulsed dynamic polarisation via periodic optimal control (\textsc{loop}) \cite{carvalho2026}. In particular, \textsc{beam}, \textsc{plato}, c\textsc{rw-opt1}, and \textsc{loop} are extremely broadband pulse sequences that were designed using advanced combinations of effective Hamiltonian methods \cite{Untidt2002,ABNielsen2019} and numerical optimal control \cite{Carvalho2025} (or non-linear optimisation).

\begin{figure}[ht!]
\centering{\includegraphics{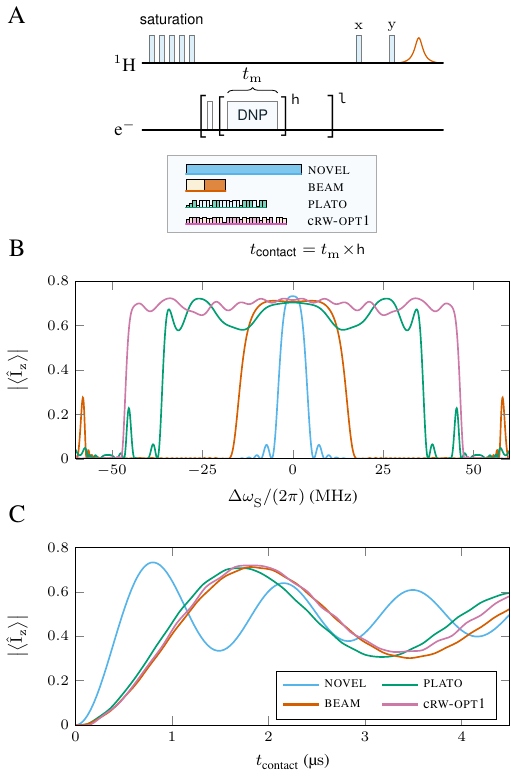}}
\caption{(A) Pulse diagrams for representative pulsed \textsc{dnp} experiments. (B,C) Numerical simulations of the $\text{e}^{-}$ to $^{1}\text{H}$ polarisation transfer efficiency $|\langle \hat{\mathrm{I}}_z\rangle|$ (numerical) as a function of the electron-spin offset $\Delta \omega_S/(2\pi)$ (represented by the \textsc{simpson} parameter \simpson{gtensor_1_iso}) (B) and the contact (or mixing) time $t_{\mathrm{contact}}$. The initial operator (\simpson{start_operator}) is $\hat{\mathrm{S}}_x$; this assumes, as in the pulse diagrams, initialisation with an ideal $(\pi/2)_y$ pulse (not included in the simulation). The detection operator (\simpson{detect_operator}) is along $\hat{\mathrm{I}}_z$, which assumes detection of transverse coherence with a solid-echo pulse sequence (not included in the simulation). The pulse sequence diagrams also contain saturation pulses and repetition of the \textsc{dnp} element to pump polarisation from the electron to multiple nuclear spins, which has not been included in the two-spin simulations. The \textsc{simpson} code and the spin-system parameters for these figures are outlined in the text and provided in the Supplementary Material.}
\label{FIG_test_bandwidth0}
\end{figure}

In the development and design of pulsed \textsc{dnp} experiments, and in their evaluation and comparison, access to numerical simulations is invaluable.  Figure~\ref{FIG_test_bandwidth0} illustrates simulations performed using \textsc{simpson} for the \textsc{novel}, \textsc{beam}, \textsc{plato}, and c\textsc{rw-opt1} pulsed \textsc{dnp} sequences. Schematics of the pulse sequences are shown in Fig.~\ref{FIG_test_bandwidth0}A, along with associated broadband excitation and \textsc{dnp} contact (or mixing) time profiles \cite{Carvalho2025,Nielsen2025,carvalho2026} at X-band MW frequencies ($14.8~\si{\mega\hertz}$ for $^{1}\text{H}$ and $9.742~\si{\giga\hertz}$ for electrons) shown in Fig.~\ref{FIG_test_bandwidth0}B and Fig.~\ref{FIG_test_bandwidth0}C, respectively. Figure~\ref{FIG_easyNMR_simview}A also shows a $2$D simulation of the offset versus contact time for the \textsc{beam} sequence, using the same parameters as in Fig.~\ref{FIG_test_bandwidth0}.

The \textsc{simpson} simulation of the \textsc{dnp} pulse sequences in Fig.~\ref{FIG_test_bandwidth0} starts with the construction of a simple \textsc{dnp} system; the Hamiltonian and spin system (the \simpson{spinsys} section) are identical to those utilised to simulate the \textsc{eseem} experiments in Fig.~\ref{FIG_eseem}. The next step is to define relevant parameters, which for the \textsc{beam} and \textsc{novel} experiments are represented by the parameters:
\begin{lstlisting}[style=SIMPSON-coloured]
par {
    proton_frequency    14.8e+6
    start_operator      I1x
    detect_operator     -I2z
    method              DNPframe
    crystal_file        rep2000
    sw                  1e9
    np                  1
    conjugate_fid       false
}
\end{lstlisting}
Again, \simpson{DNPframe} is used instead of \simpson{ROTframe} or \simpson{LABframe}. In this example, the initial operator is \simpson{I1x} and the detection operator is \simpson{-I2z}, with a powder sample specified by a \simpson{crystal_file} with 2000 \textsc{repulsion} \cite{Bak1997} crystallite angles, \simpson{rep2000}. For \textsc{novel}, the polarisation is transferred to $+\mathrm{z}$, so the detect operator should be set to \simpson{I2z} (to mitigate this, Fig.~\ref{FIG_test_bandwidth0} shows the absolute value). The spectral width \simpson{sw} and number of acquisition points \simpson{np} are also defined.

The pulse sequence in the \simpson{pulseq} section of the input file is formulated for \textsc{novel} as:
\begin{lstlisting}[style=SIMPSON-coloured]
proc pulseq {} {
    reset
    pulse   0.8 14.8e+6 180
    acq
}
\end{lstlisting}
where the pulse duration of $0.8~\si{\micro\second}$ yields maximal transfer. For \textsc{beam}, it is:
\begin{lstlisting}[style=SIMPSON-coloured]
proc pulseq {} {
    reset
    pulse   28.0e-3 32e+6   0 
    pulse   31.6e-3 32e+6 180 
    store   1
    reset
    prop    1 31
    acq 
}
\end{lstlisting}
where the two pulses of opposite phase are repeated 31 times (using a pre-calculated propagator in the \simpson{prop} command), resulting in a total duration of $1.83~\si{\micro\second}$ (corresponding to the maximum transfer efficiency for \textsc{beam} identified in Fig.~\ref{FIG_test_bandwidth0}C, \textit{vide infra}).

To examine simulation of the transfer efficiency as a function of the electron-spin offset ($\Delta \omega_\mathrm{S}^{}/(2\pi)$), the \simpson{main} section of the \textsc{simpson} input file for both \textsc{beam} and \textsc{novel} takes the form:
\begin{lstlisting}[style=SIMPSON-coloured]
proc main {} {
    set fid [open bandwidth.dat w]
    # loop over offsets
    for {set g -60e6} {$g<=60e6} {set g [expr $g+1e5]} {
        set f [fsimpson [list [list gtensor_1_iso $g]]]
        puts $fid "[format "%.0f" $g] [findex $f 1 -re]"
        funload $f
    }
    close $fid
}
\end{lstlisting}
The loop over \simpson{gtensor_1_iso} values is performed explicitly in the \simpson{main} procedure using a \textsc{tcl} \simpson{for} loop and the results are stored in the \simpson{bandwidth.dat} file.

The \simpson{pulseq} code for the \textsc{plato} and c\textsc{rw-opt1} pulse sequences is similar to that for \textsc{beam}, apart from the use of a single, convenient \simpson{pulse_shaped} command rather individual pulses:
\begin{lstlisting}[style=SIMPSON-coloured]
proc pulseq {} {
    global duration shp h
    reset
    pulse_shaped  $duration $shp
    store         1
    reset
    prop          1 $h
    acq
}
\end{lstlisting}
Here, the pulse shape \simpson{shp} is repeated \simpson{h} times, as defined in the \simpson{main} section of the \textsc{simpson} input file; for \textsc{plato}, the following lines are included:
\begin{lstlisting}[style=SIMPSON-coloured]
proc main {} {
    #...
    global duration shp h
    set shp         [load_shape plato.shp]
    set duration    [expr 0.005*[shape_len $shp]]
    set h           15
    #...
}   
\end{lstlisting}
The same structure applies to c\textsc{rw-opt1}, except with \simpson{h 13} and a different shape file, \simpson{load_shape cRW.shp}.

The code set out so far reproduces the offset plots in Fig.~\ref{FIG_test_bandwidth0}B, where it is evident that the more advanced pulse sequences lead to a significantly larger bandwidth than that for \textsc{novel}. In particular, c\textsc{rw-opt1}, designed using a combination of constrained random-walk and non-linear optimisation based on exact effective Hamiltonian theory (\textsc{eeht}) \cite{Untidt2002,Siminovitch2004}, provides a large bandwidth, here approaching $100~\si{\mega\hertz}$ \cite{Nielsen2025}. The single-spin-vector effective Hamiltonian theory (\textsc{ssv-eht}) \cite{Shankar2017,ABNielsen2019,ABNielsen2022} was a highly efficient design protocol used in the development of the \textsc{beam} \cite{Wili2022} and \textsc{plato} \cite{Nielsen2024} pulse sequences.  

Reproducing the \textsc{dnp} contact-time profiles in Fig.~\ref{FIG_test_bandwidth0}C requires modifying the \textsc{simpson} input files; this permits the simulation of data points representing an increasing number of pulse-sequence building blocks of length $\tau_\mathrm{m}$ to form the contact (or mixing) time $t_{\mathrm{contact}} = h \tau_\mathrm{m}$ under on-resonance conditions. The latter is implemented in all four examples by setting \simpson{np 5000}, thereby allocating 5000 \textsc{fid} points (a value significantly exceeding the minimum requirement) to accommodate the contact-time loop. In this case, the \simpson{main} section for \textsc{novel} and \textsc{beam} is similar to the bandwidth simulation, except there is no loop over the offsets:
\begin{lstlisting}[style=SIMPSON-coloured]
proc main {} { fsave [fsimpson] DNPtime.fid }
\end{lstlisting}
The \simpson{main} sections for \textsc{plato} and c\textsc{rw-opt1} should load the correct shape file and set the corresponding duration in the same way as for the bandwidth simulations. The \simpson{pulseq} sections differ from the bandwidth simulations, acquiring over the pulse block; for \textsc{novel} this is:
\begin{lstlisting}[style=SIMPSON-coloured]
proc pulseq {} {
    global  par
    acq_block {
        pulse 0.8 14.8e+6 180
    }   
}
\end{lstlisting}
and for \textsc{beam} the \simpson{acq_block} contains:
\begin{lstlisting}[style=SIMPSON-coloured]
    acq_block {
        pulse 28.0e-3 32e+6   0 
        pulse 31.6e-3 32e+6 180 
    }   
\end{lstlisting}
The pulse shape used for \textsc{plato} and c\textsc{rw-opt1} is implemented with \simpson{acq_block} in a similar way:
\begin{lstlisting}[style=SIMPSON-coloured]
    acq_block {
        pulse_shaped $duration $shp
    }   
\end{lstlisting}
Introduced in \textsc{simpson}\textit{-v4.0}, the \simpson{acq_block} procedure discretises the time evolution into dwell-time steps, repeating the basic element defined in parentheses as many times as necessary to populate the entire \textsc{fid} with \simpson{np} points. The offset plots in Fig.~\ref{FIG_test_bandwidth0}B were calculated using a number of repetitions, \simpson{h}, corresponding to the first maxima in the Fig.~\ref{FIG_test_bandwidth0}C \textsc{dnp} contact-time profiles.

\section{Optimal control and time-propagation}

Introducing optimal control methods for magnetic resonance can be convoluted, as they rely on idioms of control theory and numerical optimisation. To avoid alienating spectroscopists or those seeking only high-performing pulses, a simple optimal control problem is presented. The exercise presented in this section serves to introduce `fidelity' within a sterile environment and to demonstrate the analytical formula for the time-propagators of single-spin systems.

Considering a single spin, for example, a proton, with an offset $\Delta \omega=0$, the Hamiltonian for a time-dependent (shaped) pulse is written as
\begin{equation}
\mathcal{H}(t)=\omega_\mathrm{x}^{}(t)\hat{\mathrm{I}}_{\mathrm{x}}+\omega_\mathrm{y}^{}(t)\hat{\mathrm{I}}_{\mathrm{y}} \quad , \label{EQ_LinearHamiltonianHilb}
\end{equation}
where $\hat{\mathrm{I}}_{\mathrm{x},\mathrm{y}}$ are related to the Pauli matrices, $\hat{\sigma}_{\mathrm{x},\mathrm{y}}$, through Eq.~(\ref{EQ_pauli_kron}) without Kronecker products, since only a single spin is present. These operators are termed controls, and pulses can be termed control amplitudes, here denoted by $\omega_\mathrm{x,y}^{}(t)$ (in the $\mathrm{x}$ and $\mathrm{y}$ directions). Defining the \textsc{simpson} \simpson{spinsys} for this case is trivial:
\begin{lstlisting}[style=SIMPSON-coloured]
spinsys {
    channels    1H 
    nuclei      1H 
}
\end{lstlisting}

Describing the time-dependent control pulse as piecewise-constant over a small time interval $\Delta t$ \cite{Conolly1986,Mao1986}, i.e. discretising the pulse shape, the Hamiltonian of a single time-increment is written as
\begin{equation}
\mathcal{H}_n=\omega_{\mathrm{x},n}\hat{\mathrm{I}}_{\mathrm{x}}+\omega_{\mathrm{y},n}\hat{\mathrm{I}}_{\mathrm{y}} \quad .
\label{EQ_H_singlepsin1}
\end{equation}
The numerical solution to the Sch\"{o}dinger equation, using Eq.~(\ref{EQ_LinearHamiltonianHilb}), is calculated through time-ordered propagation to find a system state at a time-increment $n$ from a given initial state $\rho_0$:
\begin{equation}
\mathcal{U}_n^{}=\mathcal{P}_{\!n}^{}\dots\mathcal{P}_{\!2}^{}\mathcal{P}_{\!1}^{},\quad\rho_n=\mathcal{U}_n^{}\rho_0\mathcal{U}_n^{\dagger}  \quad .
\label{EQ_EffectivePropagatorsHilb}
\end{equation}

Typically, a propagator is calculated using a matrix exponential, $\mathcal{P}_{\!n}^{}=\exp{[-i\mathcal{H}_n\Delta t]}$; however, in the absence of additional Hamiltonian components in Eq.~(\ref{EQ_LinearHamiltonianHilb}), such as interactions, a time-propagator can be described as a rotation around the axis of the Hamiltonian $\mathcal{H}_n$ by an angle $r\Delta t$, where $r^2=\omega_{\mathrm{x},n}^2+\omega_{\mathrm{y},n}^2$. 

A time-propagator, $\mathcal{P}$, can be calculated using the Euler-Rodrigues formula \cite{Siminovitch1997} for spin-$\frac{1}{2}$:
\begin{equation}
\mathcal{P}=\begin{bmatrix} \alpha & \beta \\ -\beta^\ast & \alpha^\ast \end{bmatrix} \quad ,
\label{EQ_ER_prop_half}
\end{equation}
which is formulated in terms of the complex elements: 
\begin{equation}
\begin{gathered}
\alpha = \cos \bigg(\frac{r\Delta t}{2}\bigg)-i\frac{\omega_\mathrm{z}^2}{r} \sin \bigg(\frac{r\Delta t}{2}\bigg)\\
\beta = \frac{\omega_\mathrm{y}}{r} \sin \bigg(\frac{r\Delta t}{2}\bigg) - i\frac{\omega_\mathrm{x}}{r} \sin \bigg(\frac{r\Delta t}{2}\bigg)\\
r=\sqrt{\omega_\mathrm{x}^2+\omega_\mathrm{y}^2+\omega_\mathrm{z}^2} \quad .
\end{gathered}\label{EQ_alphabeta1}
\end{equation}
For the Hamiltonian in Eq.~(\ref{EQ_H_singlepsin1}), the term $\omega_\mathrm{z}=\Delta\omega$ is $0$, and the two controls at a time-increment therefore result in the terms $\omega_\mathrm{x}=\omega_\mathrm{x,n}$ and $\omega_\mathrm{y}=\omega_\mathrm{y,n}$. Using these variables gives the propagator at one time-increment, $\mathcal{P}_{\!n}$.

A single-spin time-propagator can also be calculated for a spin-$\frac{3}{2}$ system (which will be used in Section~\ref{Sect_quadrupole}) and is expressed in terms of $\alpha$ and $\beta$ from Eq.~(\ref{EQ_alphabeta1}) as
\begin{equation}
\mathcal{P}=
\begin{bmatrix} \alpha^3 & \sqrt{3} \alpha^2 \beta & \sqrt{3}\alpha \beta^2 & \beta^3 \\ 
-\sqrt{3}\alpha^2\beta^\ast & \alpha(\frac{3}{2}\zeta-\frac{1}{2}) & \beta (\frac{3}{2}\zeta+\frac{1}{2}) & \sqrt{3}\alpha^\ast\beta^2 \\ 
\sqrt{3}\alpha{\beta^\ast}^2 & -\beta^\ast(\frac{3}{2}\zeta+\frac{1}{2}) & \alpha^\ast(\frac{3}{2}\zeta-\frac{1}{2}) & \sqrt{3}{\alpha^\ast}^2\beta \\ 
-{\beta^\ast}^3 & \sqrt{3}\alpha^\ast{\beta^\ast}^2 & -\sqrt{3}{\alpha^\ast}^2\beta^\ast & {\alpha^\ast}^3 \end{bmatrix}\label{EQ_ER_prop_3half}
\end{equation}
with the shorthand notation $\zeta=\alpha\alpha^\ast-\beta\beta^\ast=|\alpha|^2-|\beta|^2$. Using Eq.~(\ref{EQ_ER_prop_half}) or Eq.~(\ref{EQ_ER_prop_3half}) to calculate time-propagators is much more efficient than using a matrix exponential, as no matrix operations are required. 

In the context of optimal control, a metric must be defined to be optimised: a real number that reflects the performance of a solution, known as a figure of merit. Accordingly, the aim of optimal control in magnetic resonance is to maximise the fidelity, here defined as an overlap of the final state $\rho_N^{}$ with a desired state $\rho_\text{target}$
\begin{equation}
F_{\!\mathrm{s}}=2\mathrm{Re}\big(\mathrm{tr}[\rho_\text{target}^{\dagger}\rho_N^{}]\big) \quad .
\label{EQ_fidelity_p2p}
\end{equation}
This particular fidelity, utilising a trace for the overlap of two states, is categorized as a state-to-state optimisation.

Utilising only a fidelity calculation, \textsc{simpson} can perform a gradient-free optimisation to find desired optimal pulses. A simple optimisation resulting in an inversion composite pulse to mitigate the influence of RF-field inhomogeneity is performed using the following entries in the \simpson{par} section:
\begin{lstlisting}[style=SIMPSON-coloured]
par {
    start_operator      I1z
    detect_operator     -I1z
    rfprof_file         solenoid.rf
    method              prop_split
    split_order         0

    conjugate_fid       false
    oc_optm_method      SIMPLEX
    oc_max_iter         150
}
\end{lstlisting}
Here, the \simpson{start_operator} and \simpson{detect_operator} set the control problem to find an inversion pulse. The optimisation method \simpson{SIMPLEX} (Nelder-Mead) \cite{Nelder1965} is set to run to a maximum of \simpson{150} iterations. The parameter \simpson{rfprof_file solenoid.rf} defines the static RF-inhomogeneity profile, modelling the $B_1$ field distribution in a typical solenoid coil with RF-scaling factors ranging from $40\si{\percent}$ to $100\si{\percent}$. 

The calculation method \simpson{prop_split} is requested in combination with a parameter \simpson{split_order}; this is set to zero, representing a zeroth-order approximation for the propagator-splitting method (the significance of which is clarified below). This utilises the propagator calculation in Eq.~(\ref{EQ_alphabeta1}) and ignores interactions during propagation. Since there are no interactions in Eq.~(\ref{EQ_LinearHamiltonianHilb}), the selected method is exact.

\begin{figure}
\centering{\includegraphics{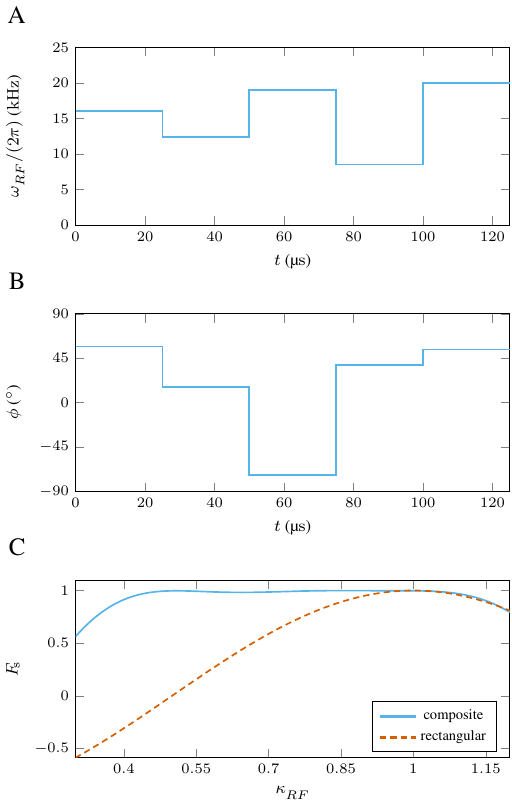}}
\caption{Gradient-free optimisation of a composite inversion pulse to mitigate large RF-field inhomogeneity. A shaped pulse with a duration of $125~\si{\micro\second}$ comprising five elements, was optimised using the \simpson{simplex} algorithm and the \simpson{prop_split} method; the resulting RF amplitudes $\omega_\mathrm{RF}/(2\pi)$ and phases $\phi$ are presented in panels (A) and (B), respectively. (C) The fidelity $F_{\!\mathrm{s}}$ of the inversion is evaluated over a range of RF-field scaling factors $\kappa_\mathrm{RF}$ and compared with the inversion profile of a rectangular pulse with a $20~\si{\kilo\hertz}$ amplitude.}
\label{FIG_simplex}
\end{figure}

The \simpson{pulseq} section propagates the shaped pulse, here named \simpson{shp}, and in this instance sets the pulse duration to $125~\si{\micro\second}$:
\begin{lstlisting}[style=SIMPSON-coloured]
proc pulseq {} {
    global shp 
    reset
    pulse_shaped 125 $shp 
    oc_acq_hermit
}
\end{lstlisting}
The internal \textsc{simpson} command \simpson{oc_acq_hermit} then calculates the fidelity in Eq.~(\ref{EQ_fidelity_p2p}). As in previous versions of \textsc{simpson}, the optimisation functionality requires a \simpson{target_function}. At the simplest level, this is formulated as:
\begin{lstlisting}[style=SIMPSON-coloured]
proc target_function {} {
    set f [fsimpson]
    set Res [findex $f 1 -re]
    funload $f
    return [format "%.20f" [expr 2.0*$Res]]
}
\end{lstlisting}
where the fidelity is multiplied by $2$, as indicated in Eq.~(\ref{EQ_fidelity_p2p}), to yield a maximum $F_{\!\mathrm{s}}=1$. The \simpson{main} section is coded as:
\begin{lstlisting}[style=SIMPSON-coloured]
proc main {} {
    global shp
    set shp [shape_create 5 -ampl 5000]
    for {set i 0} {$i < 10} {incr i 1} {
        set tfopt [oc_optimize $shp -max 20000]
    }
    save_shape $shp optimised_pulse.dat
}
\end{lstlisting}
This creates an initial pulse shape comprising $5$ piecewise-constant discrete pulses, with all initial amplitudes $\omega_{\mathrm{x},n}/(2\pi)=\omega_{\mathrm{y},n}/(2\pi)=5~\si{\kilo\hertz}$. The optimisation is then performed $10$ times, using the pulse solution from the preceding optimisation as the initial pulse for the next. This optimisation performs more effectively with a restart, as is common with gradient-free optimisations \cite{Kelley1999}. The maximal RF amplitudes in the resulting shape are restricted to $20~\si{\kilo\hertz}$, a relatively low value typical of liquid-state probes. This value was arbitrarily selected to pose a challenge for the optimisation. The resulting composite pulse is presented in Fig.~\ref{FIG_simplex}, together with its remarkable performance over a large range of RF inhomogeneity. 

\section{Propagator splitting}

It is important to explore ways to accelerate numerical calculations, particularly for large spin-systems and for the iterative determination of spin-system parameters by fitting experimental spectra. The largest part of computation time in pulse-sequence calculations is spent on the associated propagator calculation: the matrix exponential. This becomes more critical when combining spin-system simulation with numerical optimisation methods where, as outlined above, time-propagation must be repeated for many optimisation iterations. Additionally, it is common for optimal control solutions to require many discrete time-increments; more time-increments equates to more propagators. It is of great interest to consider how this propagation can be made more efficient. One method of interest in the present context, indicated recently \cite{Goodwin2022,Goodwin2023}, is splitting \cite{Trotter1959,Wilcox1967,Strang1968,Ichinose2001,McLachlan2002,Bader2014,Bader2016}. A more descriptive name, in the context of this work, is split-operator time-propagation (propagator-splitting) and an historical review of splitting methods can be found in Ref.~\cite{McLachlan2002}.

Propagator splitting offers a variety of methods to calculate a time-propagator/rotation-matrix to a desired accuracy in time/angle. One of the simplest, non-trivial, splittings is the symmetric Strang-splitting \cite{Strang1968,Marchuk1968}. Its form is instructive to the magnetic resonance community; furthermore, its derivation yields theoretical insights that release the self-imposed constraint of `memory only working backwards'.

In general, RF/MW pulse Hamiltonians $\mathcal{A}$ and interaction Hamiltonians $\mathcal{B}$ do not commute: $[\mathcal{A},\mathcal{B}]=\mathcal{A}\mathcal{B}-\mathcal{B}\mathcal{A}\neq 0$. The notation of splitting the total Hamiltonian into two parts, $\mathcal{A}$ and $\mathcal{B}$, is used in the following sections to simplify equations:
\begin{description}
\item [$\mathcal{A}$ $\Rightarrow$] Single-spin terms comprising control-pulse Hamiltonians and offset Hamiltonians. Time-propagators of these Hamiltonians can be calculated analytically with Eqs.~(\ref{EQ_ER_prop_half}) and (\ref{EQ_ER_prop_3half}).
\item [$\mathcal{B}$ $\Rightarrow$] The remainder; specifically, that which does not fit the definition of $\mathcal{A}$. Practically, this is the Hamiltonian containing all non-single-spin interactions and, in the context of optimal control, remains constant from iteration to iteration.
\end{description}
Although seemingly similar, this splitting should not be confused with that of optimal control, where the control Hamiltonian is split from the drift Hamiltonian to create a bilinear control problem \cite{GoodwinThesis}. Furthermore, many magnetic resonance simulations, aside from \textsc{mas} \textsc{nmr}, feature a time-independent $\mathcal{B}$, with the only time dependence residing in the control pulses contained in $\mathcal{A}$.

When calculating rotations or time-propagators, an exponential map exists between the Hamiltonian and the desired solution for a rotation or time-propagation, with $\mathrm{e}^{-i(\mathcal{B}+\mathcal{A})t}\neq \mathrm{e}^{-i\mathcal{B}t}\mathrm{e}^{-i\mathcal{A}t}$. A full expression for this exponential map, in which the interaction and control are split, is provided by the Zassenhaus-product formula \cite{Wilcox1967} (dropping the imaginary number and including all other common multipliers in $t$, such as the time interval $\Delta t$)
\begin{equation}
 \mathrm{e}^{(\mathcal{B}+\mathcal{A})t}\!= \mathrm{e}^{\mathcal{B}t}\mathrm{e}^{\mathcal{A}t}\mathrm{e}^{\frac{1}{2!}[\mathcal{A},\mathcal{B}]t}\mathrm{e}^{\frac{1}{3!}(2[[\mathcal{A},\mathcal{B}],\mathcal{A}]+[[\mathcal{A},\mathcal{B}],\mathcal{B}])t}\ldots
 \label{EQ_Zassenhaus}
\end{equation}
The first term, $\mathrm{e}^{\mathcal{B}t}\mathrm{e}^{\mathcal{A}t}$, is the Lie-Trotter splitting \cite{Trotter1959} which, essentially, characterizes a system where the two Hamiltonians, $\mathcal{A}$ and $\mathcal{B}$, commute; its accuracy scales linearly with $t^{-1}$. Conceptually, this is analogized with the ideal pulse in magnetic resonance, which is infinitely high and infinitesimally thin, and the approximation representing its practical implementation as a non-ideal pulse, which is generally taller than it is wide for better performance.

To account for the non-commutativity, an increase in accuracy is achieved using an integer exponent \cite{Trotter1959,Ichinose2001}
\begin{equation}
\mathrm{e}^{(\mathcal{B}+\mathcal{A})t}\!=\!\!\lim_{\eta\to\infty}\!\Big(\mathrm{e}^{\frac{1}{\eta}\mathcal{B}t} \mathrm{e}^{\frac{1}{\eta}\mathcal{A}t}\Big)^{\!\!\eta} \!\!=\! \Big(\mathrm{e}^{\frac{1}{\eta}\mathcal{B}t} \mathrm{e}^{\frac{1}{\eta}\mathcal{A}t}\Big)^{\!\!\eta} \!+ \mathcal{O}\bigg(\frac{t^2}{\eta^2}\bigg) .
 \label{EQ_LieTrotter}
\end{equation}
This is interpreted as taking the limit of the associated integral to $0$ or, in numerical terms, calculating the time-propagator for a smaller time/multiplier $\frac{t}{\eta}$ and multiplying that time-propagator by itself $\eta$ times. This is termed Trotterisation \cite{Trotter1959} and is calculated efficiently using scaling and squaring \cite{Higham2008}. \textsc{Simpson} has already implemented this functionality through the \simpson{maxdt} keyword; it is also implicit in several propagation methods, e.g. \simpson{taylor}.

As an example of simple splitting, the following broadband, single-spin optimisation is presented, in which the offset terms are split from the controls pulse terms using the Lie-Trotter splitting of Eq.~(\ref{EQ_LieTrotter}). The optimal control task aims to yield $90\si{\degree}$ broadband universal rotations by optimised pulses (\textsc{burbop}) \cite{Kobzar2012}. A simple spin system is defined, broadband $^{13}\text{C}$ over $40~\si{\kilo\hertz}$, with the Hamiltonian
\begin{equation}
\mathcal{H}(t)=\omega_\mathrm{x}^{}(t)\hat{\mathrm{I}}_{\mathrm{x}}+\omega_\mathrm{y}^{}(t)\hat{\mathrm{I}}_{\mathrm{y}}+\omega_\text{\textsc{I}}^{}\hat{\mathrm{I}}_{\mathrm{z}} \label{EQ_LinearHamiltonianHilb_off}
\end{equation}
which is implemented in \textsc{simpson} as:
\begin{lstlisting}[style=SIMPSON-coloured]
spinsys {
    channels    13C
    nuclei      13C
    shift       1 1 0 0 0 0 0
}

par {
    averaging_file      shift_1_iso_40kHz.ave
    method              prop_split
    split_order         1
    conjugate_fid       false
    
    oc_max_iter         500
    oc_cg_min_step      1e-4
}
\end{lstlisting}
An \simpson{averaging_file} averages the optimisation over a list of chemical shifts contained in \simpson{shift_1_iso_40kHz.ave}. This file contains $\omega_\mathrm{I}^{}/(2\pi)\in[-20,+20]~\si{\kilo\hertz}$ as a grid with $41$ equally weighted elements, with weights normalised to unity (as before, a nominal chemical shift is set to $1~\si{\hertz}$, which is overwritten at every stage of the simulation with the \simpson{averaging_file}).

The Lie-Trotter splitting of Eq.~(\ref{EQ_LieTrotter}) is implemented using \simpson{split_order 1}. \textsc{Simpson} currently calculates pulses using Eq.~(\ref{EQ_ER_prop_half}) (i.e. using $\omega_\mathrm{z}^{}=0$ in Eq.~(\ref{EQ_alphabeta1})) for the single-spin Hamiltonian $\mathcal{A}$. This corresponds to setting $\omega_\mathrm{I}=0$ for nuclei or $\omega_\mathrm{S}=0$ for electrons and allocating these offsets to the interaction Hamiltonian in $\mathcal{B}$.

In a departure from the standard \textsc{burbop}, a new class of pulse is identified using target propagators dispersed around the transverse plane to achieve higher fidelity. This is a relatively new approach in optimal control and was first published as second-order phase dispersion by optimised rotation (\textsc{sordor}) pulses \cite{Goodwin2020}. This work demonstrated that the duration of optimised pulses is halved when compared to \textsc{burbop} without loss of performance. \textsc{Sordor} pulses were subsequently validated in experiments \cite{Haller2022,Joseph2023}.

The dispersion of the target propagators, as detailed in Ref.~\cite{Goodwin2020}, is given by an angle $\alpha_k$ defined as
\begin{equation}
\alpha_k = \pi b Q \bigg( 1 - \frac{\omega_\mathrm{I,k}^2}{\Omega^2} \bigg) ,
\label{EQ_sordor_ak}
\end{equation}
where $Q$ is the quadratic coefficient, $Q\in[0,1]$, $\omega_\mathrm{I,k}$ is a particular offset within a range of $[-\Omega/2,+\Omega/2]$, and $b$ is the bandwidth factor, $b=\Omega t$ ($t$ is the duration of the pulse). Calculation of $\alpha_k$ requires the current offset (chemical shift, $\omega_\mathrm{I,k}$); this is achieved in \textsc{simpson} by inspecting the current Hamiltonian matrix (rounding is used to set $\omega_\mathrm{I,k}$ to whole numbers of hertz): 
\begin{lstlisting}[style=SIMPSON-coloured]
proc get_dsp {Omega duration Q} {
    set pi  3.14159265358979323846
    set b   [expr $duration*$Omega]
    set H   [matrix get hamiltonian]
    set h11 [expr round([lindex $H 0 0 0]/($pi))]
    set ak  [expr $pi*$b*$Q*(1-(($h11**2)/($Omega**2)))]
    return $ak
}
\end{lstlisting}

In optimisations of universal rotations, fidelity is defined as the overlap of the effective propagator $\mathcal{U}_N^{}$ with the desired effective propagator $\mathcal{U}_\text{target}$. In this particular set-up, it is formulated as
\begin{equation}
F_{\!\mathrm{u}}=\frac{1}{4}\mathrm{Re}\big(\mathrm{tr}[\mathcal{U}_\text{target}^{\dagger}\mathcal{U}_N^{}]\big) \quad .
\label{EQ_fidelity_ur}
\end{equation}
The factor of $\frac{1}{4}$ normalises the maximum fidelity to unity.

The desired effective propagator, 
\begin{equation}
\mathcal{U}_\text{target} = \exp \big[-i \frac{\pi}{2} (\hat{\mathrm{I}}_{\mathrm{x}}\cos\alpha_k + \hat{\mathrm{I}}_{\mathrm{y}}\sin\alpha_k ) \big], 
\label{EQ_sordor_utarget}
\end{equation}
can be determined using \simpson{avgham_static} and stored in a specific memory slot for use in \simpson{oc_acq_prop} calculating the overlap of the propagators. The \simpson{pulseq} is formulated as:
\begin{lstlisting}[style=SIMPSON-coloured]
proc pulseq {} {
    global shp Omega duration Q
    
    set ak [get_dsp $Omega $duration $Q]
    set cosak [expr cos($ak)]
    set sinak [expr sin($ak)]
    reset
    avgham_static 0.25e6 $cosak*(I1x)+$sinak*(I1y)
    store 10
    reset
    pulse_shaped $duration $shp
    oc_acq_prop 10
}
\end{lstlisting}
In this instance, the rotation angle of $\pi/2$ is set with the first argument of \simpson{avgham_static} representing the duration of action (in $\si{\micro\second}$) of the Hamiltonian corresponding to the second argument. The amplitude of the given Hamiltonian (the axis in the transverse plane, \simpson{I1x} and \simpson{I1y}) is $1~\si{\radian}/\si{\second}$ and thus the duration is \simpson{0.25e6}. 

The \simpson{target_function} is identical to that in the previous example, except normalised with \simpson{Res/4.0}, i.e. divided by the square of the matrix dimension. In addition to the \simpson{target_function}, this optimisation requires a gradient-following optimisation method to make progress. The \textsc{simpson} \simpson{gradient} function is formulated as:
\begin{lstlisting}[style=SIMPSON-coloured]
proc gradient {} {
    global par NOC
    set par(np) $NOC
    set g [fsimpson]
    return $g
}
\end{lstlisting}
This requires \simpson{par(np)} to be set to the number of points in the pulse shape, \simpson{NOC}, to accommodate all gradient elements in \simpson{g}.

Following the procedure detailed in Ref.~\cite{Goodwin2020}, a morphic optimal control task commences from a known pulse, \textsc{burbop}, and progressively increases the quadratic coefficient of the phase dispersion to determine higher fidelity \textsc{sordor} pulses. This is implemented in the \simpson{main} section as:
\begin{lstlisting}[style=SIMPSON-coloured]
proc main {} {
    global shp NOC Omega Q b duration
    
    # load the BURBOP shape   
    set shp [load_shape burbop090_0300us.dat]
    set NOC [shape_len $shp]
    set duration [expr $NOC/2]
    set Omega 40e3
    
    for {set Q 0.0} {$Q<=1.00} {set Q [expr $Q+0.01]} {
        set tfopt [oc_optimize_phase $shp -max 10000]
    }
}
\end{lstlisting}
This optimisation incorporates a new functionality in \textsc{simpson}, optimising only the phase of the pulse, as is appropriate for \textsc{sordor}, using the keyword \simpson{oc_optimize_phase}. The tolerance for an acceptable gradient is reduced from the default (\simpson{1e-3}) to \simpson{oc_cg_min_step 1e-4}, consistent with the original \textsc{sordor} study \cite{Goodwin2020}.

\begin{figure}
\centering{\includegraphics{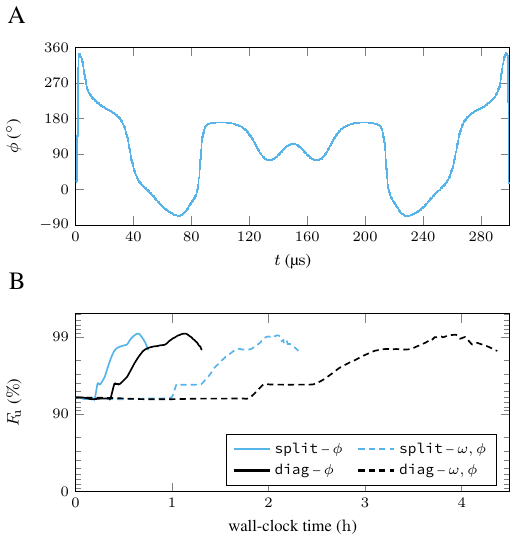}}
\caption{(A) Plot of the constant-amplitude ($10~\si{\kilo\hertz}$) \textsc{sordor} pulse \cite{Goodwin2020}, with $Q=0.84$ and $F_{\!\mathrm{u}}=99.095\si{\percent}$, optimised using the \textsc{simpson} code outlined in the text. (B) Wall-clock time for fidelities achieved with morphic optimal control, starting from \textsc{burbop} \cite{Kobzar2012} and ramping the quadratic coefficient, $Q$ (\textsc{burbop} has $Q=0$), for both \simpson{method diag} and \simpson{method prop_split}; dashed lines represent an optimisation of phase and amplitude, while solid lines represent phase-only optimisation. Wall-clock time were measured using a Linux workstation with an \textsc{amd} Ryzen 7 5700G 8-core processor.}
\label{FIG_sordor}
\end{figure}

Results are illustrated in Fig.~\ref{FIG_sordor}, with the optimal pulse shown in Fig.~\ref{FIG_sordor}A; this was identified at $Q=0.84$ with a fidelity of $F_{\!\mathrm{u}}=99.095\si{\percent}$ using a phase-only optimisation (at a constant amplitude of $10~\si{\kilo\hertz}$), which is consistent with the original study \cite{Goodwin2020}. The maximum fidelity achieved at each $Q$ is shown in Fig.~\ref{FIG_sordor}B, plot as a function of the wall-clock time. The four plot line provide a comparison between \simpson{method diag} and \simpson{method prop_split}; the solid lines represent the phase-only optimisation \simpson{oc_optimize_phase} (performed as in the original \textsc{sordor} paper and subsequent research \cite{Goodwin2020,Haller2022,Buchanan2025}), while the dashed lines represent a simultaneous phase and amplitude optimisation \simpson{oc_optimize} (which results in an approximately constant-amplitude pulse with slightly less fidelity).

\subsection{High accuracy splittings}

The splittings described above only treat single-spin terms effectively: the Hamiltonians $\mathcal{H}_{\mathrm{x}}$, $\mathcal{H}_{\mathrm{y}}$, and $\mathcal{H}_{\mathrm{z}}$. To simulate interactions between spins effectively with splitting methods, higher accuracy is required.

Taking the matrix logarithm of the Zassenhaus formula \cite{Wilcox1967} in Eq.~(\ref{EQ_Zassenhaus}) leads to the Baker-Campbell-Hausdorff (\textsc{bch}) formula \cite{Baker1905,Campbell1897,Hausdorff1906}, which is familiar to the \textsc{nmr} community. For $\mathrm{e}^{\mathcal{C}}= \mathrm{e}^{\mathcal{B}}\mathrm{e}^{\mathcal{A}}$, this is
\begin{equation}
\mathcal{C}= \mathcal{B}+\mathcal{A}+{\frac{1}{2!}[\mathcal{A},\mathcal{B}]}+{\frac{1}{3!}(\mathcal{A}_2)}\ldots \quad ,
\end{equation}
where the shorthand notation $\mathcal{A}_2=2[[\mathcal{A},\mathcal{B}],\mathcal{A}]+[[\mathcal{A},\mathcal{B}],\mathcal{B}]$ is employed. 

An additional order of accuracy can be obtained, relatively cheaply when compared to Eq.~(\ref{EQ_LieTrotter}), through evaluation at the midpoint of $\mathcal{B}$ in Eq.~(\ref{EQ_Zassenhaus}); this is analogous to the increased accuracy of an integral midpoint evaluation \cite{Vinding2021} and yields the symmetrized Zassenhaus formula \cite{Bader2014,Bader2016}
\begin{equation}
 \mathrm{e}^{(\mathcal{B}+\mathcal{A})t}\!=\ldots \mathrm{e}^{\frac{1}{2}\frac{1}{4!}(\mathcal{A}_2)t^3_{}}\mathrm{e}^{\frac{1}{2}\mathcal{B}t}\mathrm{e}^{\mathcal{A}t}\mathrm{e}^{\frac{1}{2}\mathcal{B}t}\mathrm{e}^{\frac{1}{2}\frac{1}{4!}(\mathcal{A}_2)t^3_{}}\ldots
\end{equation}
The first (central) term is named the symmetric Strang-splitting \cite{Strang1968,Marchuk1968}, which is accurate to $t^2$
\begin{equation}
\mathrm{e}^{(\mathcal{B}+\mathcal{A})t}\!=\mathrm{e}^{\frac{1}{2}\mathcal{B}t} \mathrm{e}^{\mathcal{A}t} \mathrm{e}^{\frac{1}{2}\mathcal{B}t}+\mathcal{O}(t^3) \quad .\label{EQ_strang_splitting}
\end{equation}
The result of this symmetrization is to gain an order of accuracy without additional costly matrix exponentials, requiring only an ancillary matrix-matrix multiplication.

Higher-order splittings, typically of even order when based upon the symmetric Zassenhaus formula, are obtained from a mathematically rich field employing order conditions and the theory of composition of numerical integrators \cite{McLachlan1995,Murua1999,Blanes2008} (which is beyond the scope of this manuscript). Without detailing the specifics of higher-order propagator-splittings, their general form is a conjugate-palindromic chain of matrix exponential multiplications, each with a multiplier coefficient. In the case of Strang-splitting, the coefficients are $1$ multiplied by $\mathcal{A}$ and $\frac{1}{2}$ multiplied by $\mathcal{B}$. The palindromic form is evident for Strang-splitting by noting the central symmetry in Eq.~(\ref{EQ_strang_splitting}).

Based upon the symmetric Zassenhaus form, a general, even-order, palindromic splitting comprises additional coefficients
\begin{equation}
\mathrm{e}^{(\mathcal{B}+\mathcal{A})t}\!=\!\cdots\mathrm{e}^{b_2^{}\mathcal{B}t} \mathrm{e}^{a_2^{}\mathcal{A}t}\mathrm{e}^{b_1^{}\mathcal{B}t} \mathrm{e}^{a_1^{}\mathcal{A}t} \mathrm{e}^{b_1^{}\mathcal{B}t}\mathrm{e}^{a_2^{}\mathcal{A}t}\mathrm{e}^{b_2^{}\mathcal{B}t}\!\cdots
\end{equation}
The coefficients $a_i$ and $b_i$ may be positive or negative; negative coefficients are interpreted as propagation backwards in time and, as detailed in \cite{Blanes2005}, splitting orders $>2$ must incorporate some negative coefficients. Splitting of this type increases accuracy at the cost of additional matrix operations.

Scrutinising the Strang-splitting and Lie-Trotter splitting further, where $a_1=1$, one of the order conditions shows how the two are numerically related: $\sum_ib_i=1$. Recasting the Strang-splitting more generally to include asymmetric forms
\begin{equation}
\mathrm{e}^{(\mathcal{B}+\mathcal{A})t}\!=\mathrm{e}^{a_1^{}\mathcal{B}t} \mathrm{e}^{\mathcal{A}t} \mathrm{e}^{a_2^{}\mathcal{B}t}+\mathcal{O}(t^\kappa) \quad ,
\label{EQ_GeneralSplit}
\end{equation}
where $\kappa\in[2,0, 3.0]$ specifies the error term and $a_1^{}+a_2^{}=1$ is an order condition. Any values for $a_1^{}$ and $a_2^{}$ may be chosen according to $a_1^{}+a_2^{}=1$; the limit of $a_1^{}=1$ and $a_2^{}=0$ yields the first term of Eq.~(\ref{EQ_Zassenhaus}) (Lie-Trotter splitting), which possessed an error term in Eq.~(\ref{EQ_GeneralSplit}) with $\kappa=2$. The optimum in this context is $\kappa=3$ with $a_1^{}=a_2^{}=\frac{1}{2}$.

For simulations with small $t$, such as those in optimal control, $4^\text{th}$-order splitting \cite{Blanes2002} is potentially too accurate, whereas $2^\text{nd}$-order is insufficiently precise. Furthermore, utilising splitting for adaptive optimal control (not described here) \cite{Goodwin2022} performs more effectively with gradual increases in accuracy (e.g. from $2^\text{nd}$ to $3^\text{rd}$ to $4^\text{th}$ orders) rather than a sharp jump from $2^\text{nd}$-order to $4^\text{th}$-order. Recent publications provide some odd-orders of accuracy, $3^\text{rd}$-order \cite{Blanes2022} and $5^\text{th}$-order \cite{Bernier2023}, but these are less intuitive; they are conjugate-palindromic. As an example, the $3^\text{rd}$-order splitting \cite{Blanes2022} is
\begin{equation}
\mathrm{e}^{(\mathcal{B}+\mathcal{A})t}\!=\mathrm{e}^{a_{1}^{\ast}\mathcal{B}t} \mathrm{e}^{b_{1}^{}\mathcal{A}t} \mathrm{e}^{a_{2}^{\ast}\mathcal{B}t} \mathrm{e}^{b_{2}^{}\mathcal{A}t} \mathrm{e}^{a_{2}^{}\mathcal{B}t} \mathrm{e}^{b_{1}^{}\mathcal{A}t} \mathrm{e}^{a_{1}^{}\mathcal{B}t}+\mathcal{O}(t^4),
\end{equation}
where, interestingly, the coefficients $a_i$ are complex and $b_i$ are real. The interpretation of complex coefficients is left to the reader. In addition to the splitting orders $0$, $1$, $2$, $3$, $4$, and $5$, \textsc{simpson} also incorporates a $6^\text{th}$-order splitting \cite{Omelyan2003} for very high accuracy. Although no specific syntax is provided to define the Trotter-number $\eta$ in Eq.~(\ref{EQ_LieTrotter}), \textsc{simpson} effectively implements this through \simpson{maxdt}.

To benchmark the splitting methods, the numerical simulation of the $\text{e}^{-}$ to $^{1}\text{H}$ polarisation transfer efficiency as a function of the electron spin offset (Fig.~\ref{FIG_test_bandwidth0}B) was evaluated at every $1~\si{\hertz}$. Specifically, the \textsc{plato} pulse sequence utilising \simpson{method diag} was calculated in approximately $67~\si{\second}$ using a Linux workstation equipped with an Intel Core i5-3210M processor. The same simulation was performed using \simpson{method prop_split}, resulting in the following wall-clock times: \simpson{split_order 2}, $19~\si{\second}$; \simpson{split_order 3}, $33~\si{\second}$; \simpson{split_order 4}, $58~\si{\second}$; \simpson{split_order 5}, $80~\si{\second}$; \simpson{split_order 6}, $92~\si{\second}$.

The polarisation transfer efficiency values, $|\langle \hat{\mathrm{I}}_z\rangle|$, obtained via \simpson{split_order 2} are inadequate, as the time increment is too large for this approximation. In contrast, \simpson{split_order 3} is sufficiently accurate, while \simpson{split_order 4} becomes indistinguishable from the exact \simpson{method diag}. Decreasing \simpson{maxdt} improves results; by defining \simpson{maxdt 0.025} ($\eta=2$), the wall-clock time for \simpson{split_order 2} is $27~\si{\second}$ and for \simpson{split_order 3} is $60~\si{\second}$. Decreasing the value further to \simpson{maxdt 0.0125} ($\eta=4$), the wall-clock time for \simpson{split_order 2} becomes $42~\si{\second}$. Results of the polarisation transfer efficiency values, $|\langle \hat{\mathrm{I}}_z\rangle|$, for these simulations are supplied in the Supplementary Material.

\section{Pulse Transients}

In the detailed evaluation of magnetic resonance experiments, it is often relevant to include further details concerning experimental parameters. This includes RF and/or MW inhomogeneity, amplitude-digitisation and time-digitisation of pulses, and the response function of pulses. This aspect has been demonstrated clearly by composite pulses and optimal control pulses, which are resilient to field inhomogeneities. Furthermore, pulse/phase transients have been considered in numerous studies \cite{Mehring1972,Barbara1991,Wittmann2015,Rasulov2025}. Compared with typical cases, mitigating pulse transients may be even more important for pulsed \textsc{epr} and \textsc{dnp} \cite{Spindler2012}, where the MW pulses seen by a sample may deviate significant from those transmitted from the console. A transfer matrix may be utilised to neutralise this effect: by adding an antenna to the resonator \cite{Spindler2013,Probst2019}; by employing a sample with a narrow \textsc{epr} line \cite{Kaufmann2013}; through phase variation in nutation frequency experiments \cite{Doll2013}; via feedback control through a gradient-free optimisation \cite{Goodwin2018,Verstraete2022}; with an analytical model of the distortions \cite{Iriarte2025}; or by using a calculated response function \cite{Rasulov2025}.

To address pulse transients relating to magnetic resonance in general terms (i.e. \textsc{nmr}, \textsc{epr}, and \textsc{dnp}), an implementation of pulse-response functions is provided in \textsc{simpson}\textit{-v6.0}. This is founded upon the central premise that a pulse shape may be optimised in a relatively coarse manner, while the dynamics are calculated with a significantly finer discretisation by following the coil-distorted shape \cite{Borneman2012,Spindler2013,Hincks2015,Wittmann2015,Probst2019}. Although a fidelity calculation requires only this, a gradient-following optimisation also necessitates a transform of the gradient back from the fine discretisation of the spin dynamics to the coarse discretisation of the optimiser.

\begin{figure}
\centering{\includegraphics{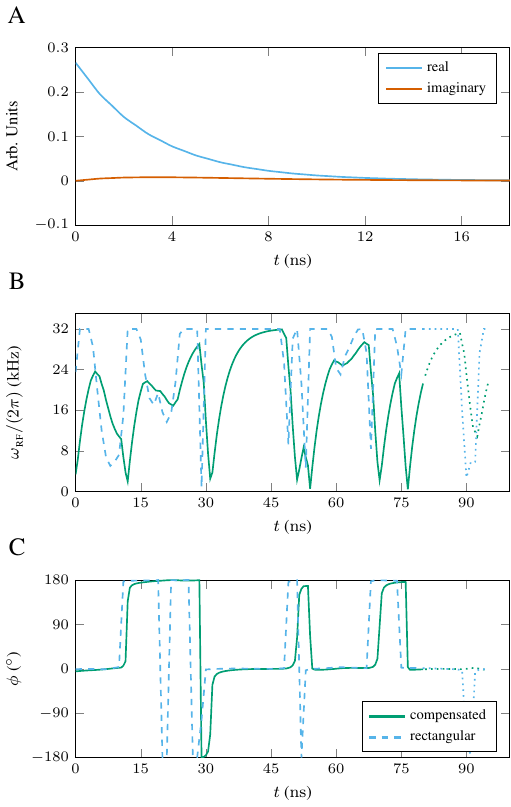}}
\caption{(A) Impulse-response function $h(t)$, comprising $40$ discrete points in intervals of $0.5~\si{\nano\second}$, utilised for optimal control of a broadband \textsc{dnp} experiment with amplitudes $\omega_\mathrm{RF}/(2\pi)$ (B) and phases $\phi$ (C) for the initial (staring guess) of the optimised \textsc{dnp} experiment. (B,C) Dashed, blue lines illustrate the piecewise-constant rectangular pulses, whereas solid, green lines illustrate the transient-compensated distorted pulses.}
\label{FIG_test_responsefun}
\end{figure}

The mathematical formalism is established by defining a discretised distortion operator, $\varphi^{}_{nm}$, and an impulse-response function, $h(t)$. For simplicity in the derivation, only a single irradiation channel and a single control shape are considered. To avoid confusion with non-distorted pulses, an angular frequency $\upsilon$ (covering both phase components) is utilised rather than the $\omega_{x,y}$ components employed above. Conceptually, the irradiation that the spin `feels' (transient-compensated), $\upsilon(t)$ (lower-case, smaller time interval, henceforth $\delta t$), is a distorted form of the input shape (rectangular, piecewise-constant, pulses), $\Upsilon(t)$ (upper-case, larger time interval, henceforth $\Delta t$), and is modelled as a convolution with an impulse response (theoretical or experimentally measured), $h(t)$
\begin{equation}
\upsilon(t)=\Upsilon(t)\ast h(t)=\!\!\int\limits_{-\infty}^{+\infty}\!\!\!\Upsilon(t_{}^{\prime})h(t-t_{}^{\prime})\mathrm{d}t_{}^{\prime} \quad .
\end{equation}
Henceforth, distorted pulses $\upsilon(t)$ are referred to as \textit{transient-compensated pulses}, whereas input pulses that have not undergone convolution are referred to as \textit{rectangular pulses}. An example of an impulse-response function is shown in Fig.~\ref{FIG_test_responsefun}A.

Since $\Upsilon(t)$ is piecewise-constant and is only physically defined for $t>0$, the integral is rewritten as a discrete sum, with $\upsilon(t)$ becoming
\begin{equation}
\upsilon(t)=\sum\limits_{n=1}^{N}\Upsilon_{n}^{}\!\!\!\!\!\!\int\limits_{(n-1)\Delta t}^{n\Delta t}\!\!\!\!\!\!h(t-t_{}^{\prime})\mathrm{d}t_{}^{\prime} \quad .
\end{equation}
Evaluating at the centre of a small interval $\delta t$, the discrete transient-compensated pulse may be defined as
\begin{equation}
\upsilon^{}_{m}:=\upsilon\Big(\big(m-\tfrac{1}{2}\big)\delta t\Big)=\sum\limits_{n=1}^{N}\Upsilon_{n}^{}\!\!\!\!\!\!\int\limits_{(n-1)\Delta t}^{n\Delta t}\!\!\!\!\!\!h\Big(\big(m-\tfrac{1}{2}\big)\delta t-t_{}^{\prime}\Big)\mathrm{d}t_{}^{\prime} \quad .
\end{equation}
The integral part of this equation defines the required distortion operator, which is also cast in a discrete form
\begin{equation}
\varphi_{mn}^{}=\!\!\!\!\!\!\int\limits_{(n-1)\Delta t}^{n\Delta t}\!\!\!\!\!\!h\Big(\big(m-\tfrac{1}{2}\big)\delta t-t_{}^{\prime}\Big)\mathrm{d}t_{}^{\prime}=\!\!\!\!\!\!\!\!\sum\limits_{k=(n-1)r+1}^{nr}\!\!\!\!\!\!\!\!h^{}_{m-k+1}\delta t \quad .
\label{EQ_dstn_oper}
\end{equation}
The positive integer $r\in \mathds{Z}^{+}$, where $r=\Delta t/\delta t$, requires $\Delta t$ to be a multiple of $\delta t$ in \textsc{simpson}. This formalises the distortion operator as a matrix of size $N\times M$, multiplied by an input vector of length $N$ to result in an output vector of length $M$, where $M>N$ and $M/N\in\mathds{Z}^{+}$.

At a distant time, the transients vanish and the pulse is characterized by the nominal values of pulse amplitude and phase, i.e. a long, continuous pulse converges to the desired input pulse amplitude and phase. Consequently, the convolution at distant times yields a normalisation
\begin{equation}
\upsilon(t\gg0)=A=\!\!\int\limits_{-\infty}^{+\infty}\!\!\!Ah(t-t_{}^{\prime})\mathrm{d}t_{}^{\prime}\,\,\,\,\,\Rightarrow\,\,\, \int\limits_{-\infty}^{+\infty}\!\!\!h(t_{}^{\prime})\mathrm{d}t_{}^{\prime}=1 \quad ,
\end{equation}
giving a discrete, complex normalisation factor
\begin{equation}
C=\sum\limits_{k=1}^{M}h^{}_{k} \quad .
\end{equation}

\subsection{Optimal control with pulse transients}

Returning to the subject of optimal control, the fidelity gradient requires a convolution to transform the distorted physical-frame (that of the transient-compensated pulse) back to the computational-frame (that of the rectangular pulse). Although various methods exist to define fidelity, its essential property is a measure of the performance of a pulse in executing a desired task; namely, it serves as a figure of merit \cite{GoodwinThesis}.

One such fidelity definition, as in Eq.~(\ref{EQ_fidelity_p2p}), is the overlap of the final system state (calculated through time-ordered propagation of individual piecewise-constant rectangular pulses from a given initial state $\rho_0^{}$) with a desired state $\rho_\text{target}^{}$. The gradient of the fidelity $\nabla F_{\!\mathrm{s}}$ requires the derivatives of each of these time-propagators with respect to the control at that time-increment; specifically, the discrete rectangular pulse. With the inclusion of pulse transients, the chain rule yields the required propagator derivative as
\begin{equation}
\frac{\partial \mathcal{U}}{\partial \Upsilon_n^{}}=\sum\limits_{m=1}^{M}\frac{\partial \mathcal{U}}{\partial \upsilon_m^{}}\frac{\partial \upsilon_m^{}}{\partial \Upsilon_n^{}} \quad .
\label{EQ_deriv_jacob}
\end{equation}
The derivatives of the distorted transient-compensated pulse propagators are determined using the gradient-ascent pulse engineering \textsc{grape} optimal control method \cite{Khaneja2005,deFouquieres2011,Goodwin2016,Goodwin2023}
\begin{equation}
\frac{\partial \mathcal{U}}{\partial \upsilon_m^{}} = \mathcal{P}_{\!M}^{}\cdots \mathcal{P}_{\!m+2}^{} \mathcal{P}_{\!m+1}^{} \frac{\partial \mathcal{P}_{\!m}^{}}{\partial \upsilon_m^{}}  \mathcal{P}_{\!m-1}^{}\mathcal{P}_{\!m-2}^{}\cdots \mathcal{P}_{\!0}^{} \quad.
\end{equation}

The Jacobian in Eq.~(\ref{EQ_deriv_jacob}) may be conveniently calculated using the distortion operator defined in Eq.~(\ref{EQ_dstn_oper}), with
\begin{equation}
\frac{\partial \upsilon_m^{}}{\partial \Upsilon_n^{}} = \frac{\partial}{\partial \Upsilon_n^{}}\Bigg(\sum\limits_{k=1}^{N}\varphi_{mk}^{}\Upsilon_k^{} \Bigg) = \varphi_{nm}^{} \quad .
\end{equation}
This provides the necessary gradient elements with respect to the rectangular pulses (the coarse, non-distorted, input pulse) via a simple multiplication of the transient-compensated pulse propagator derivatives by the transpose of the distortion operator
\begin{equation}
\frac{\partial \mathcal{U}}{\partial \Upsilon_n^{}} = \sum\limits_{m=1}^{M}\big[\varphi_{mn}^{}\big]^{\!\mathrm{T}}\frac{\partial \mathcal{U}}{\partial \upsilon_m^{}} \quad .
\label{EQ_grad_trans}
\end{equation}

To demonstrate the implementation of pulse-transients in \textsc{simpson}, the following simple electron-nuclear two-spin system is created in \simpson{spinsys}:
\begin{lstlisting}[style=SIMPSON-coloured]
spinsys {
    channels    e
    nuclei      e 1H
    gtensor     1 1 0 0 0 0 0
    hyperfine   1 2 0 0.8676e+6 0 60 0
}
\end{lstlisting}
This spin-system comprises an electron and $^{1}\text{H}$ nucleus, with control-pulses applied to the electron channel. The $g$-tensor is isotropic and, as previously, is set to a nominal value of $1~\si{\hertz}$ to permit this parameter to be varied in an \simpson{averaging_file}. The hyperfine interaction possesses no isotropic component, an anisotropy of $0.8676~\si{\mega\hertz}$, and a Euler angle $\beta_\mathrm{PL}$ of $60\si{\degree}$. The Hamiltonian for this system is equivalent to that of Eq.~(\ref{EQ_spinsys1}) with $A\approx -0.22~\si{\mega\hertz}$ and $B\approx 1.13~\si{\mega\hertz}$.  

Common parameters, including those specific to optimal control, are set in the \simpson{par} section as:
\begin{lstlisting}[style=SIMPSON-coloured]
par {
    proton_frequency    14.8e+6
    crystal_file        alpha0beta0
    gamma_angles        1
    sw                  1e9
    averaging_file      gtensor_1_iso_pm70MHz.ave
    start_operator      I1x
    detect_operator     -I2z
    method              DNPframe diag
    
    # Parameters for optimisation
    conjugate_fid       false
    oc_grad_level       2
    oc_max_iter         1000
    oc_tol_cg           1e-9
}
\end{lstlisting}
This simulation configured to optimise a state-to-state problem for offsets $\Delta \omega_\mathrm{S}^{}/(2\pi)\in\{-70,0,+70\}~\si{\mega\hertz}$, contained in the provided \simpson{averaging_file}, for a static, single-crystal spin-system defined through \simpson{gamma_angles} and \simpson{crystal_file}, respectively. In this instance, the time-propagation method is set to \simpson{diag} to provide a reference for subsequent methods.  It should be emphasized that the optimisation was set up for a single-crystal system rather than a powder, as typically used for state-to-state optimal control in solid-state \textsc{nmr} \cite{Kehlet2004,Kehlet2007,Tosner2009}. This choice was inspired by the recent finding that for periodic \textsc{dnp} pulse sequences optimised using a state-to-state-like figure-of-merit function, it could be proved \cite{Carvalho2025a} that single-crystal optimisation with a finite pseudo-secular coupling yielded the same optimal offset profile as a powder, subject to a scaling factor \cite{Nielsen2024,Nielsen2025}. In this context, it is noted that the chosen finite (high) value of the pseudo-secular coupling $B$, required for \textsc{dnp} without irradiation on the nuclear spins, was ensured by $B=\frac{3}{2}b_{IS}\sin{2\beta_\mathrm{PL}}$ using $\beta_\mathrm{PL}=60\si{\degree}$ as set in \simpson{spinsys}.

The initial stage in utilising pulse-transients resides in the \simpson{main} part of the \textsc{simpson} input file; specifically, the timings are compatible with \textsc{dnp} optimal control applications: 
\begin{lstlisting}[style=SIMPSON-coloured]
proc main {} {
    global shp phi shpdist duration
   
    # time resolution is 1 ns (each element is 1 ns)
    set t 0.001
    set shp [load_shape shape_oc_initial.dat]
    set N [shape_len $shp]
    set duration [expr {$N*$t}]
  
    # distortion operator, time resolution 0.5ns
    set dt [expr {$t/2.0}]
    set phi [create_distortion_operator Rsp.dat $dt $N $t] 
    
    # allocate distorted shape with M elements
    set M [expr {int($N*$t/$dt)}]
    set shpdist [shape_create $M]

    # optimisation
    set tfopt [oc_optimize $shp -max 32e6]
    save_shape $shp oc_tansients_final.dat
  
    free_distortion_operator $phi
    free_all_shapes
}
\end{lstlisting}
In the example above, a file containing a pulse-response function, \simpson{Rsp.dat}, is prepared in advance, in addition to an initial pulse guess, \simpson{shape_oc_initial.dat}. The initial pulse guess was produced from a separate, non-distorted optimisation assuming $1.0~\si{\nano\second}$ elements constituting the total shape (not shown here). The response function for this example is illustrated in Fig.~\ref{FIG_test_responsefun}A, assuming time resolution of $0.5~\si{\nano\second}$. The distortion operator $\varphi_{mn}^{}$ in Eq.~(\ref{EQ_dstn_oper}) is implemented with \simpson{create_distortion_operator}, with an associated \simpson{free_distortion_operator} at the end of the input file (see Table~\ref{tab_commands}). 

A distinct \simpson{phi} (and associated response function) is required for each control-pulse channel subject to distortion. Significantly, the flexibility to utilise different pulse-response functions for various irradiation channels is advantageous, e.g. in the optimisation or simulation of heteronuclear solid-state \textsc{nmr} experiments where different circuit $Q$-factors apply at distinct frequencies. 

The \simpson{pulseq} section propagates the distorted shaped pulse (here named \simpson{shpdist}):
\begin{lstlisting}[style=SIMPSON-coloured]
proc pulseq {} {
    global shpdist duration
    reset
    pulse_shaped $duration $shpdist
    oc_acq_hermit
}
\end{lstlisting}
The internal \textsc{simpson} command \simpson{oc_acq_hermit} then calculates the fidelity defined in Eq.~(\ref{EQ_fidelity_p2p}). The application of the distortion within the \simpson{target_function} is formulated as:
\begin{lstlisting}[style=SIMPSON-coloured]
proc target_function {} {
    global par shp phi shpdist

    # create distorted shape
    distort_shape $phi $shp $shpdist
    
    # calculate target (one number)
    set par(np) 1
    set f [fsimpson]
    set Resn [expr [findex $f 1 -re] ]
    funload $f
    
    return [format "%.20f" $Resn]
}
\end{lstlisting}
where the shape \simpson{shp} is distorted to an output \simpson{shpdist} using the \simpson{distort_shape} function and the previously created distortion operator \simpson{phi}. The remainder of the \simpson{target_function} comprises standard optimal control functionality. Note that the distorted shape must be regenerated before every call to \simpson{fsimpson}, as the optimisation procedure iteratively updates the original shape \simpson{shp}.

The \simpson{gradient} function for the \simpson{target_function} above necessitates further scrutiny:
\begin{lstlisting}[style=SIMPSON-coloured]
proc gradient {} {
    global par shp phi shpdist
   
    # create distorted shape, prepare gradient
    distort_shape $phi $shp $shpdist
    oc_grad_shapes $shpdist
    set par(np) [shape_len $shpdist]
    set g [fsimpson]
   
    # reconstruct grads w.r.t. original shapes
    set gg [fcreate -np [shape_len $shp] -sw $par(sw)]
    reconstruct_gradient $g $gg $phi
    funload $g

    return $gg
}
\end{lstlisting}
Specifically, an initial step involves distorting the required shape with \simpson{distort_shape}, consistent with the \simpson{target_function}. A gradient with respect to the pulse parameters of the distorted shape is initialised with \simpson{par(np)} points. The output gradient required by the optimiser is then generated with a number of points corresponding to the original shape; this is utilised in the \simpson{reconstruct_gradient} function, according to Eq.~(\ref{EQ_grad_trans}), using the distortion operator \simpson{phi} (the necessary transpose of which is calculated internally by \textsc{simpson}).

\begin{figure*}
\centering{\includegraphics{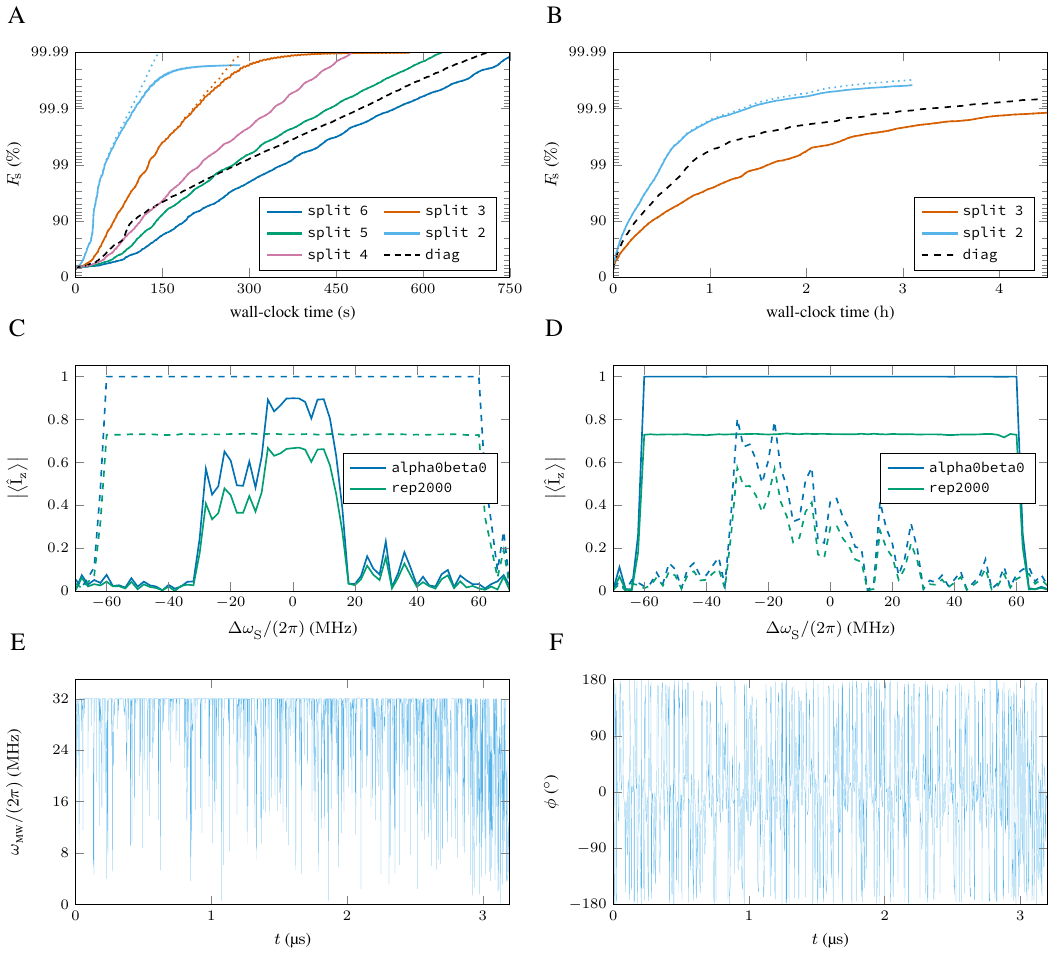}}
\caption{(A) Simulation time for three simultaneous state-to-state optimisations covering offsets $\Delta \omega_\mathrm{S}^{}/(2\pi)\in\{-70,0,+70\}~\si{\mega\hertz}$. Dotted lines represent the fidelity determined via the splitting approximation, while solid lines represent the same data points determined using an exact fidelity calculation (the exact fidelity calculation is omitted from the timing). (B) Simulation time for an ensemble of state-to-state optimisations covering offsets $\Delta \omega_\mathrm{S}^{}/(2\pi)\in[-60,+60]~\si{\mega\hertz}$ with 61 equally space offsets. (C) Bandwidth of the initial pulse utilised in (B), simulated with \simpson{crystal_file rep2000} (green) and the single-crystal \simpson{crystal_file alpha0beta0} (blue). Dashed lines denote the simulation excluding transients, whereas solid lines incorporates transients. (D) Bandwidth of the final pulse produced in (B), simulated with \simpson{crystal_file rep2000} (green) and the single-crystal \simpson{crystal_file alpha0beta0} (blue). Dashed lines denote the simulation excluding transients, whereas solid lines incorporates transients. The final pulse that produces the blue, solid lines in (D) is illustrated in (E) for amplitude and (F) for phase. Wall-clock time was measured using a Linux workstation with an \textsc{amd} Ryzen 7 5700G 8-core processor.}
\label{FIG_test_bandwidth_transient}
\end{figure*}

The convergence of the fidelity is illustrated for \simpson{split_order} $2$, $3$, $4$, $5$, $6$, and \simpson{diag} as a function of wall-clock time for a single-crystal in Fig.~\ref{FIG_test_bandwidth_transient}A. This optimal control problem is somewhat synthetic, comprising three simultaneous state-to-state problems, and is intended primarily to demonstrate the acceleration in comparison to the examples in the previous splitting study \cite{Goodwin2022}.

When considering a broadband ensemble, the concept of \textit{exact optimisation on an approximate landscape} \cite{Jensen2021,Goodwin2022} loses a little applicability: the ensemble constitutes an average optimisation, and the average smooths the approximation within the optimisation landscape. This is evident in Fig.~\ref{FIG_test_bandwidth_transient}B for an ensemble $\Delta \omega_\mathrm{S}^{}/(2\pi)\in[-60,+60]~\si{\mega\hertz}$ with 61 equally space offsets. Here, \simpson{split_order 2} remains the most rapid method, whereas \simpson{split_order 3} is hampered by an approximation that is too accurate for certain members of the ensemble. Conversely, \simpson{split_order 2} does not yield fidelities that diverge from an exact calculation of the fidelity (the dotted and solid lines in Fig.~\ref{FIG_test_bandwidth_transient}B are almost identical). Essentially, the error in the fidelity calculation is averaged away. This finding indicates that \simpson{split_order 2} is sufficient for ensemble optimisation and, in the presented example, gives a simulation that requires approximately half the time of an exact fidelity calculation (\simpson{diag} in this instance).

Figures ~\ref{FIG_test_bandwidth_transient}C and D illustrate offset profiles simulated using \simpson{crystal_file rep2000} (green) and as a single crystal (blue). The dashed lines represent simulations excluding transients (rectangular pulse), while the solid lines incorporate transients (transient-compensated pulses) generated with the response function in Fig.~\ref{FIG_test_responsefun}. Figure ~\ref{FIG_test_bandwidth_transient}C depicts the bandwidth of the initial pulse before optimisation, and Fig.~\ref{FIG_test_bandwidth_transient}D depicts the bandwidth of the pulse produced from the optimisation that utilises the transients from the response function. Evidentially, the optimisation commences from a pulse that is optimal in the absence of pulse transients. The amplitude and phase of the pulse generated with the transient-inclusive optimisation are shown in Figs.~\ref{FIG_test_bandwidth_transient}E and F, respectively. For all simulations, it should be noted that the offset profiles for the single-crystal and powder simulations are virtually identical, to within a scaling factor of $0.73$. Despite this being the onset for the optimisation, it is remarkable that this remains the case for state-to-state optimisation of pulse shapes with very long duration. This significantly accelerates optimal control optimisations, as demonstrated earlier in the optimisation of periodic \textsc{dnp} pulse sequences \cite{Nielsen2024,Nielsen2025}, and may serve as inspiration for state-to-state optimal control optimisations for powder samples in more general terms. In the context of optimisation of powder \textsc{dnp} pulse sequence by single-crystal optimisation, it should be mentioned that periodic single-crystal optimised \textsc{dnp} pulse sequences can, remarkably, be translated directly into powder \textsc{mas} solid-state \textsc{nmr} dipolar recoupling sequences \cite{Carvalho2025a}. 

\section{Quadrupolar second-order cross-term interactions}\label{Sect_quadrupole}

Another area of considerable and increasing interest within the solid-state \textsc{nmr} community concerns quadrupolar spin nuclei, which will finalise examples of novel features in this version of \textsc{simpson}. This implementation highlights the necessity for the continuous development of the software platform to address  challenges in high-order simulations, as increasingly advanced experiments on progressively challenging samples are performed. 

Quadrupolar nuclei, possessing a spin quantum number larger than $\frac{1}{2}$, play an important role in materials research, with a growing interest in the characterization of energy-storage materials and functional materials. Frequently, the quadrupolar coupling interaction is strong and must be treated using higher-order perturbation theory. For the pure quadrupolar coupling interaction, such features are already implemented in \textsc{simpson} and controlled via the \simpson{quadrupole} keyword, whereby the user specifies the desired perturbation level, ranging from $1$ (first order) to $3$ (third order).
There are, however, other important higher-order terms, such as second-order contributions that mix the quadrupolar coupling interaction with either the chemical-shielding anisotropy (\textsc{csa}) or the dipolar coupling interactions of the quadrupolar nucleus. A detailed theoretical description of these second-order Hamiltonian cross-terms is provided, for example, in the work of Ashbrook et al. \cite{Ashbrook2009} and is not repeated here. The new features presented here concern simulation of residual dipolar splitting \cite{Harris1992} and $2$D \textsc{stmas} experiments \cite{Zhehong2000}. 

\begin{figure}
\centering{\includegraphics{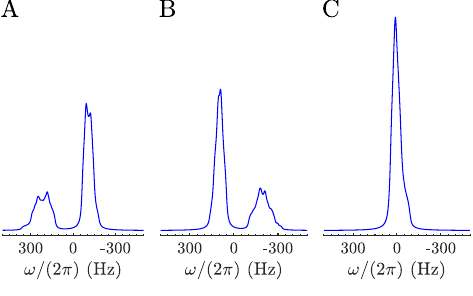}}
\caption{$^{1}\text{H}$ \textsc{mas} \textsc{nmr} spectra demonstrating the residual dipolar splitting effect of a dipole-dipole coupled $^{14}\text{N}$ nucleus with a strong quadrupolar interaction. The actual $^{1}\text{H}$ line-shape depends on the mutual orientation of the dipolar and quadrupolar tensors, expressed using the Euler angles $\alpha$, $\beta$, $\gamma$ that define the orientation of the dipolar tensor within the quadrupolar principal-axis frame. In (A), the two tensors are assumed to be collinear, i.e. $\alpha=0\si{\degree}$, $\beta=0\si{\degree}$, $\gamma=0\si{\degree}$, whereas in (B) and (C), $\beta=90\si{\degree}$, $\gamma=0\si{\degree}$ and $\beta=90\si{\degree}$, $\gamma=90\si{\degree}$ are specified, respectively. Other parameters include: $C_Q=3~\si{\mega\hertz}$, $b_\mathrm{IS}/(2\pi) = -7~\si{\kilo\hertz}$, a \textsc{mas} frequency of $50~\si{\kilo\hertz}$, and a $^{1}\text{H}$ Larmor frequency of $800~\si{\mega\hertz}$.}
\label{FIG_rds}
\end{figure}

Residual dipolar splitting is observed when a spin-$\frac{1}{2}$ nucleus, typically a high-$\gamma$ spin such as $^{1}\text{H}$, is dipole-dipole coupled with a quadrupolar nucleus possessing a large quadrupolar coupling, e.g. $^{14}\text{N}$. The consequence of such interactions is the splitting of the resonance line into two components with an intensity ratio of $2:1$ and a complicated powder pattern that depends upon the mutual orientation of the two interaction tensors, as demonstrated in Fig.~\ref{FIG_rds}. Depending on the coupling strength, this effect may challenge spectral resolution in compounds where nitrogen atoms are not isotopically labelled with $^{15}\text{N}$. Such detrimental broadening may be reduced/alleviated by heteronuclear decoupling of the $^{14}\text{N}$ spin. However, this is not a straightforward task, given the general difficulty in developing efficient and broadband decoupling sequences for quadrupolar spin nuclei \cite{Nehra2025}. In such instances, numerical simulation may provide essential insight into the decoupling performance \cite{Nehra2026}.  

\begin{figure*}[ht!]
\centering{\includegraphics{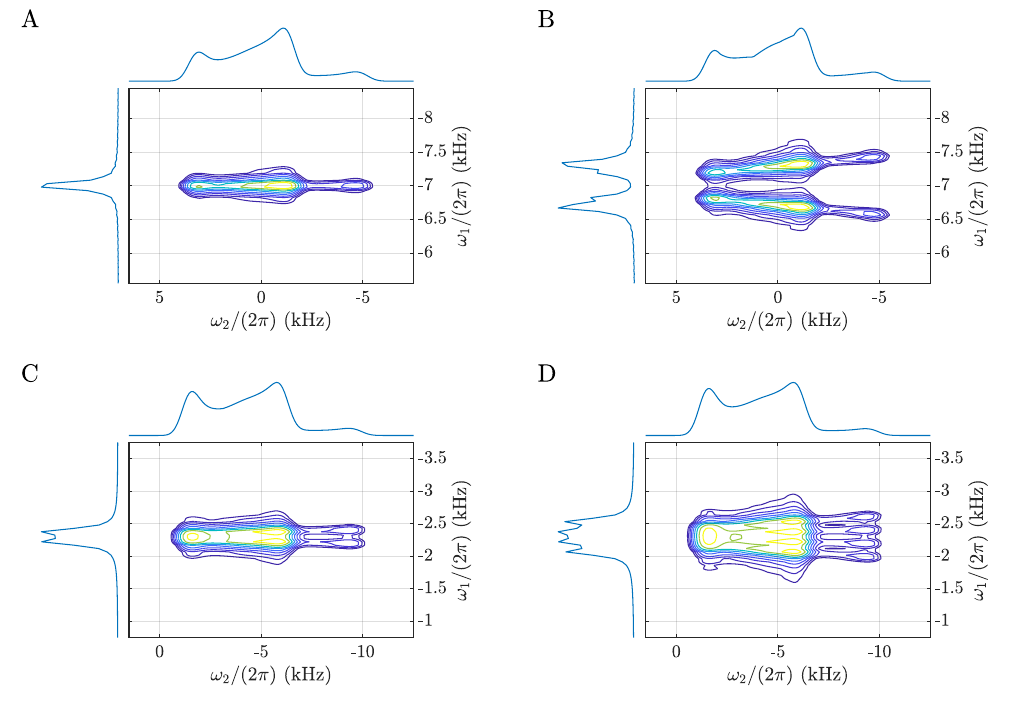}}
\caption{$2$D \textsc{stmas} spectra (split-$t1$ variant) calculated assuming a spin $\mathrm{I}=\frac{3}{2}$ nucleus ($100~\si{\mega\hertz}$ Larmor frequency, $50~\si{\kilo\hertz}$ \textsc{mas} frequency, $4~\si{\mega\hertz}$ quadrupolar-coupling constant, and zero asymmetry) and real pulses ($100~\si{\kilo\hertz}$ RF amplitude with $1.6~\si{\micro\second}$ pulse length for satellite-transition excitation and conversion to the central-transition; $10~\si{\kilo\hertz}$ amplitude and $25~\si{\micro\second}$ length for the soft central-transition inversion). (A) Simple pattern is obtained when cross-term effects are ignored. (B) The peak splits under the influence of the quadrupolar–\textsc{csa} cross-term (\textsc{csa} of $500~\si{ppm}$, zero asymmetry). (C) and (D) illustrate the effect of the cross-term between the quadrupolar and dipolar interactions (dipole-dipole coupling constant of $10~\si{\kilo\hertz}$) with a spin $\mathrm{I}=\frac{1}{2}$ and a spin $\mathrm{I}=\frac{3}{2}$ nucleus, respectively. The interaction tensors are assumed collinear in all cases.}
\label{FIG_stmas}
\end{figure*}

Another substantial effect of the quadrupolar second-order cross-terms on spectral line-shapes may be identified in $2$D \textsc{stmas} experiments. In such experiments, finite RF pulses may be utilised to excite and convert satellite transitions to the central transition, employing a split-$t1$ variant of the experiment. Representative simulated spectra are illustrated in Fig.~\ref{FIG_stmas}. The second-order cross-term between the quadrupolar coupling and the anisotropic shielding (\textsc{csa}) interactions of a spin-$\frac{3}{2}$ nucleus leads to a separation of the two satellite transitions. This separation, depending on the crystallite orientation, is not refocussed in the \textsc{stmas} experiment. Similarly, the $2$D peak pattern is split into two or four components when the quadrupolar spin-$\frac{3}{2}$ is dipole-dipole coupled to another spin-$\frac{1}{2}$ or spin-$\frac{3}{2}$ spin, respectively.

A logical prerequisite for such simulations in \textsc{simpson} is the definition of \simpson{quadrupole}, \simpson{shift}, and/or \simpson{dipole} interactions within the \simpson{spinsys} section of a \textsc{simpson} input file. Additional Hamiltonians, describing second-order cross-terms are triggered individually for each cross-term with the syntax:
\begin{lstlisting}[style=SIMPSON-coloured]
    quadrupole_x_shift N1
    quadrupole_x_dipole N1 N2
\end{lstlisting}
where \simpson{N1} is the index of the quadrupolar nucleus, as defined within the \simpson{nuclei} keyword, and \simpson{N2} is another nucleus dipole-dipole coupled to the quadrupolar nucleus \simpson{N1}. When generating the cross-term Hamiltonians (indicated by \simpson{x}), the parameters of the corresponding coupling tensors (\simpson{quadrupole}, \simpson{shift}, \simpson{dipole}) are utilised automatically; this includes their relative orientations, as defined by the Euler angles of their \textsc{pas} within the crystal-fixed frame.

Simulations presented in Fig.~\ref{FIG_stmas} provide an informative example for analysing calculation speed and comparing different methods implemented in \textsc{simpson}. A quadrupolar nucleus with spin-$\frac{3}{2}$ is assumed, for which the Hamiltonian matrix dimension is $4\times4$ in Hilbert space. The quadrupole and \textsc{csa} interactions, and their cross-term, are all described by diagonal operators $\mathrm{I}_\mathrm{z}$; the calculation of the necessary propagators is trivial, implemented with an analytical integration over the time period during sample rotation as the default method in \textsc{simpson} \cite{Bak2000}. The most time-consuming part of simulation is the computation of the RF-pulse propagator, where the total Hamiltonian is no longer diagonal. By default, \textsc{simpson} utilises matrix diagonalization via eigen-decomposition (\simpson{method diag}). The spectra in Figs.~\ref{FIG_stmas}A and B were calculated in approximately $277~\si{\second}$ using a Linux workstation with the \textsc{amd} Ryzen 9 7950X 16-core processor. This work utilises the splitting method, which splits the total Hamiltonian into interactions (a diagonal matrix in this case) and RF pulses, for which the analytical formula is employed (see Eq.~(\ref{EQ_ER_prop_3half})). Utilising \simpson{method prop_split} and \simpson{split_order 2}, the calculation took $267~\si{\second}$, resulting in a negligible speed-up due to the small matrices involved in the computation. Using \simpson{split_order 3} leads to a significantly longer wall-clock time of $760~\si{\second}$, as more operations are required for propagator construction at higher precision. Note, however, that all three calculation methods yield equivalent results, with visually indistinguishable $2$D spectra.

When the quadrupole nucleus is dipolar-coupled to a spin-$\frac{1}{2}$ (the case illustrated in  Fig.~\ref{FIG_stmas}C), the Hamiltonian matrix dimension is $8\times8$ in Hilbert space. The interaction Hamiltonian is diagonal (heteronuclear dipole-dipole interaction is truncated to the $\mathrm{I}_\mathrm{z} \mathrm{S}_\mathrm{z}$ term) and the RF Hamiltonian is off-diagonal. Simulations using the default diagonalization method required $1619~\si{\second}$ in this instance due to the increased matrix dimensions. In a previous version of \textsc{simpson} \cite{Tosner2014}, the \simpson{block_diag} method was introduced; this is utilised used in cases when the RF field is applied only on one channel (the quadupolar nucleus) while the coupled coupled spin-$\frac{1}{2}$ nucleus remains unperturbed. Under these conditions, the RF Hamiltonian is decomposed into two diagonal blocks of size $4\times4$, and the eigen-decomposition is applied twice to each smaller matrix. The calculation is accelerated, finishing in $548~\si{\second}$ wall-clock time, almost three-times faster. 

The \simpson{prop_split} method avoids diagonalization altogether; calculations with \simpson{split_order} set to $2$ or $3$ were completed in $276~\si{\second}$ and $802~\si{\second}$, respectively. In the final case (Fig.~\ref{FIG_stmas}D), the quadrupolar nucleus is dipolar-coupled to a heteronuclear spin-$\frac{3}{2}$, resulting in Hamilatonian matrix dimension of $16\times16$ in Hilbert space. The full diagonalization (the default) and the block-diagonalization (\simpson{method block_diag}, with the Hamiltonian divided into four $4\times4$ blocks) required $3556~\si{\second}$ and $2547~\si{\second}$, respectively. Using the approximate splitting method with \simpson{split_order} $2$ or $3$, the corresponding wall-clock times were $287~\si{\second}$ and $860~\si{\second}$, respectively, providing an order of magnitude acceleration in this particular case. The wall-clock times for the splitting method remain relatively constant across all cases presented in Fig.~\ref{FIG_stmas}, increasing only moderately with the Hamiltonian dimensions. The success of \simpson{split_order 2} arises from calculating propagators over short time-increments (set to $0.05~\si{\micro\second}$ via the \simpson{maxdt} parameter) in order to follow closely time-modulations of the strong quadrupolar coupling interaction ($4~\si{\mega\hertz}$) induced by sample rotation at a \textsc{mas} frequency of $50~\si{\kilo\hertz}$. Corresponding input files are provided in Supplementary Material.
 
\section{Conclusion}

With focus on the strongly increasing needs for advanced numerical software enabling design, understanding, and interpretation of advanced experiments in magnetic resonance, we have in this work introduced a substantially upgraded version of \textsc{simpson} including numerous new features with particular focus on the experiments also involving electron spins. This immediately supports calculation of pulsed \textsc{epr} and \textsc{dnp} experiments. To further comply with increasingly advanced pulse sequences and larger spin systems in real experimental settings we have also introduced novel aspects of optimal control, advanced propagator splitting, \textsc{mas} modulated field inhomogeneities, and handling of pulse transients. We have also addressed enhanced user and supplementary software interaction for visualisation. With this we have addressed new needs for efficient simulation software, in particular embracing the rapidly developing interface between \textsc{nmr}, \textsc{epr}, and \textsc{dnp}. Realising that this may only represent part of the state-of-today needs, but certainly not all needs also including in future developments in magnetic resonance, we have also changed the programming fundament from C in the current versions of \textsc{simpson} to \CC{} in a gitlab repository with increased level of documentation for community development of future versions of \textsc{simpson}.

\begin{acknowledgments}
The authors acknowledge support from the Villum Foundation Synergy programme (grant 50099), the Novo Nordisk Foundation (NERD grant NNF22OC0076002), Horizon 2020 (grant 101008500), and the DeiC National HPC (g.a. DeiC-AU-N5-2024094-H2-2024-35). Z.T. acknowledges support from the Czech Science Foundation, project No. 24-13437L.
\end{acknowledgments}

\section{Data availability}

All input files and simulated data are available from Zenodo (xxx). In addition, \textsc{e}asy\textsc{nmr} workflows for each simulation is available from \url{https://easynmr.csdm.dk/simpson5}.

\bibliography{Simpson6_refs}

\end{document}